\documentclass[%
 aip,
 amsmath,amssymb,
 reprint,%
nofootinbib    
]{revtex4-1}

\usepackage{graphicx}
\usepackage{dcolumn}
\usepackage{bm}
\usepackage{caption}    
\usepackage{subcaption}    
\usepackage{array}  
\usepackage[hidelinks,pagebackref,colorlinks=true,allcolors=blue]{hyperref}	

\usepackage{xcolor}   
\usepackage{soul}   
\usepackage[export]{adjustbox}  

\usepackage[utf8]{inputenc}
\usepackage[T1]{fontenc}
\usepackage{mathptmx}
\usepackage{etoolbox}
\usepackage{environ}         

\DeclareUnicodeCharacter{2060}{!!!!I-AM-HERE!!!!!}   
\DeclareUnicodeCharacter{03BD}{!!!!I-AM-HERE!!!!!}   
\DeclareUnicodeCharacter{03C3}{!!!!I-AM-HERE!!!!!}   

\makeatletter
\def\@email#1#2{%
 \endgroup
 \patchcmd{\titleblock@produce}
  {\frontmatter@RRAPformat}
  {\frontmatter@RRAPformat{\produce@RRAP{*#1\href{mailto:#2}{#2}}}\frontmatter@RRAPformat}
  {}{}
}%
\makeatother

\makeatletter
\newcommand{\@citeX}[3]{{%
  \@ifundefined{b@#2}{\mbox{\reset@font\bfseries ?}%
    \G@refundefinedtrue
    \@latex@warning
      {Citation `#2' on page \thepage \space undefined}}%
  {#1\csname b@#2\endcsname#3}}}
\newcommand{\citeA}[1]{\@citeX{[Ref.~}{#1}{]}}
\newcommand{\citeB}[1]{\@citeX{Reference~}{#1}{}}
\newcommand{\citeC}[1]{\@citeX{}{#1}{}}
\makeatother

\newlength{\myl}
\let\origequation=\equation
\let\origendequation=\endequation

\RenewEnviron{equation}{
  \settowidth{\myl}{$\BODY$}                       
  \origequation
  \ifdimcomp{\the\linewidth}{>}{\the\myl}
  {\ensuremath{\BODY}}                             
  {\resizebox{\linewidth}{!}{\ensuremath{\BODY}}}  
  \origendequation
}

\begin{document}

\title{Effect of detachment on Magnum-PSI ELM-like pulses: I. Direct observations and qualitative results}

\author{Fabio Federici}

\altaffiliation{E-mail address: fabio.federici@ukaea.uk}
\affiliation{ 
York Plasma Institute, Department of Physics, University of York, Heslington, York, YO10 5DD, United Kingdom
}%
\affiliation{ 
Oak Ridge National Laboratory, Oak Ridge, Tennessee 37831, USA
}%
\affiliation{ 
United Kingdom Atomic Energy Authority, Culham Centre for Fusion Energy, Culham Science Centre, Abingdon, Oxon, OX14 3DB, United Kingdom
}%

\author{Bruce Lipschultz}%
\affiliation{ 
York Plasma Institute, Department of Physics, University of York, Heslington, York, YO10 5DD, United Kingdom
}%

\author{Gijs R. A. Akkermans}%
\affiliation{ 
DIFFER-Dutch Institute for Fundamental Energy Research, De Zaale 20, 5612 AJ Eindhoven, The Netherlands
}%

\author{Kevin Verhaegh}%
\affiliation{ 
United Kingdom Atomic Energy Authority, Culham Centre for Fusion Energy, Culham Science Centre, Abingdon, Oxon, OX14 3DB, United Kingdom
}%
\affiliation{ 
York Plasma Institute, Department of Physics, University of York, Heslington, York, YO10 5DD, United Kingdom
}%

\author{Matthew L. Reinke}%
\affiliation{ 
Oak Ridge National Laboratory, Oak Ridge, Tennessee 37831, USA
}%
\affiliation{ 
Commonwealth Fusion Systems, Cambridge, MA 02139, USA
}%



\author{Ivo G. J. Classen}%
\affiliation{ 
DIFFER-Dutch Institute for Fundamental Energy Research, De Zaale 20, 5612 AJ Eindhoven, The Netherlands
}%

\author{Magnum-PSI Team}%
\affiliation{ 
DIFFER-Dutch Institute for Fundamental Energy Research, De Zaale 20, 5612 AJ Eindhoven, The Netherlands
}%


\begin{abstract}
Conditions similar to those at the end of the divertor leg in a tokamak were replicated in the linear plasma machine Magnum-PSI. The neutral pressure in the target chamber is then increased to cause the target to transition from an attached to a detached state. Superimposed to this steady state regime, ELM-like pulses are reproduced, resulting in a sudden increase in plasma temperature and density, such that the heat flux increases transiently by half an order of magnitude. 
Visible light emission, target thermography, and Thomson scattering are used to demonstrate that the higher the neutral pressure the more energy is removed from the ELM-like pulse in the volume. If the neutral pressure is sufficiently high, the ELM-like pulse can be prevented from affecting the target and the plasma energy is fully dissipated in the volume instead (ID 4 in \autoref{tab:table1}). The visible light images allow the division of the ELM-plasma interaction process of ELM energy dissipation into 3 "stages" ranging from no dissipation to full dissipation (the target plasma is detached).
In the second publication related to this study, spectroscopic data is analysed with a Bayesian approach, to acquire insights into the significance of molecular processes in dissipating the plasma energy and particles.

\end{abstract}

\maketitle


\section{Introduction}\label{introduction}

In tokamaks the core exhaust is directed to a small portion of the divertor target through the Scrape Off Layer (SOL). The parallel heat flux entering the divertor region in a power plant size reactor is more than 2 orders of magnitude higher of what materials can withstand with current technologies. Thus mitigation methods are required to reduce the surface perpendicular heat flux to below the widely used limit of $10MW/m^2$. \cite{Pitts2019} One such mitigation, planned for usage in ITER, is detachment; a buffer plasma region with low temperature is created in the divertor such that a large fraction of the plasma energy and particles are dissipated (via radiation and various recombination processes) rather than reaching the surface. 

Detachment is induced using several methods: increasing the density upstream of the divertor, seeding impurities or reducing the power flow leaving the core plasma and entering the divertor. Impurities can retain electrons in multiple shells for a wider temperature range than hydrogen, radiating more efficiently in the SOL and divertor, reducing the power flowing along field lines. Such radiative losses lead to a temperature reduction such that volumetric processes (recombination and charge exchange) become relevant. Increasing the core edge density leads to an increase in the divertor plasma density, as well as of that of hydrogen neutrals that can remove energy radiatively or via transport, again cooling the divertor plasma. \cite{Leonard2018} In these low temperature and high density cases a region with strong emissivity “detaches” from the target and moves upstream towards the x-point. Detachment was demonstrated in Alcator C-Mod \cite{Lipschultz2007} and AUG \cite{Kallenbach2015a} to significantly reduce target heat flux in conditions scalable to burning plasmas like ITER.

In high core performance regimes short (sub ms) bursts of heat and particles from the core (edge localised modes, ELMs) happen cyclically increasing temporarily the heat flux by 2-3 orders of magnitude, and this can hardly be tolerated by large tokamaks. \cite{Jachmich2011} If ELMs happen when the target is detached, the divertor plasma temporarily reattaches and looses energy in the process of dissociating and ionising the neutrals. This is beneficial for the reduction of target heat load but computational and experimental investigations are difficult due to the dynamic nature of the phenomena.

The aim of this study is to investigate the behaviour of ELM-like pulses on a detached target plasma and to understand the role of various processes in removing energy from the ELM before it reaches the target.
Another topic is to determine if a regime in which the ELM energy can effectively be fully dissipated in the volume, meaning the energy reaching the target is negligible for its whole duration, exists. 
In this first paper on the topic it is shown what can be inferred from direct measurements of the plasma properties (temperature, density, light emission) and interaction with the target (thermography).

To reliably investigate the above phenomena, ELM-like pulses are produced in Magnum-PSI, a linear plasma machine in the DIFFER laboratory, the Netherlands. Magnum PSI is capable of producing steady state target perpendicular heat fluxes comparable to that at at the ITER target. The configuration is different than a tokamak, but the simpler geometry allows for an easier interpretation, better repeatability and diagnostic access. The ELM-like pulses are generated thanks to a capacitor bank (CB) connected in parallel to the steady state plasma source and detachment is induced by increasing hydrogen neutral pressure in the target chamber.

Linear machines can be used to study the tokamak exhaust, but have significant differences. In a tokamak divertor, before and during detachment and the high recycling regime especially, the majority of the ions that contribute to the target flux originate from the ionisation of neutrals recycling in the divertor, and the upstream acts as the source of energy for said ionisation. This in virtue of a plasma temperature before detachment $>10eV$. 
In linear machines instead, due to the lower plasma temperature (typically up to $5-10eV$), the ionisation source in the volume is usually much smaller than the ion flux from the plasma source upstream, hence recycling at the target is not important.

Another major difference is the connection length: tens of meters in tokamaks compared to about half a meter in Magnum-PSI. While in Magnum-PSI the transit time from source to target is almost negligible, in a tokamak the wave of hotter plasma can have a complex interaction with the cold neutrals buffer generated by detachment. Linear machines are therefore more representative of a similar cross field transport than of the ELM burn through. Additionally, in tokamaks ELMs can assume a filamentary 3D structure that may change as they move along field lines, further complicating their interaction with the steady-state plasma and neutrals.

The ELM frequency is dictated by the type of regime, and influences various aspects like the energy released per ELM\cite{Jachmich2011} and potentially the neutral pressure\cite{Smith2020}. In a linear machine the ELMs are released in a controlled way, so different interactions can be studied independently.

Impurities likely also have a significant influence on the power, momentum and particle during ELMs, and this could too be different in linear machines.
Despite these differences, linear machines remain valuable tools for understanding the processes occurring during tokamak ELMs and the effects of neutral pressure.

In this paper it will be shown that by increasing the target chamber neutral pressure, energy is removed from the ELM-like pulses. In some cases, the target is not significantly affected by the ELM-like pulses. The pulse effect on the target is comparable to what is measured in current tokamaks, even if significantly lower than the expectation for large scale devices like ITER, and can be reduced by increasing the neutral pressure.


\section{Experimental setup}\label{Experiments setup}

\begin{figure*}
	\centering
	\includegraphics[width=\linewidth,trim={30 0 0 0},clip]{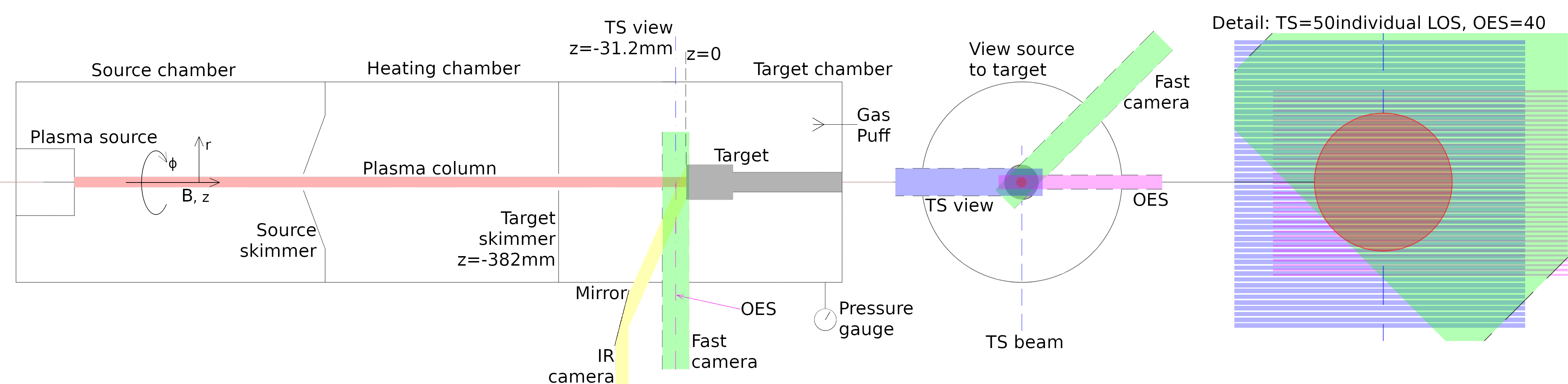}
	\caption{Schematic of Magnum-PSI linear plasma machine. $z=0$ corresponds to the surface of the target, Thomson scattering (TS, in purple) and Optical Emission Spectroscopy (OES, in pink) are located at $z=-31.2$mm. The TS view is composed of 50 evenly spaced Lines of Sight (LOS) orthogonal to the plasma column length ($z$). The OES view is composed of 40 LOS perpendicular to the column length. The fast camera, for measuring integrated light emission from the plasma, has a radial view of the plasma (in green) through a lateral viewport of the vacuum vessel of diameter $\sim$75mm. The IR camera has a radial view (in yellow) that is converted to looking at the target surface by a mirror outside the target chamber. The mirror angle can be adapted to maintain the view of the target when moved in $z$.}
	\label{fig:layout}
\end{figure*}

Magnum-PSI is a linear machine with superconducting coils surrounding the vacuum chamber to confine the plasma generated by a cascaded-arc source from it to the target. The vacuum vessel is split in three chambers separated by skimmers to allow differential pumping. This ensures that the conditions in the target chamber are independent from the source and the vast majority of the impurities and neutrals from the source are removed from the plasma column before it enters the target chamber. \cite{Scholten2013} Magnum-PSI is capable of reproducing ITER steady state orthogonal target heat flux, although the parallel heat flux is much lower. \cite{Scholten2013} A capacitor bank is connected in parallel to the plasma source power supply to temporarily increase the heat flux tenfold and simulate ELMs. \cite{Morgan2014} A schematic of the Magnum-PSI machine is in \autoref{fig:layout}.

\subsection{Experimental conditions}\label{Experimental conditions}

\autoref{tab:table1} indicates the experimental conditions used on Magnum-PSI for this work, with the detachment stage of the discharge based on the visible light emission defined in \autoref{Fast camera}. ID 1-4 are referred to as weak pulses: the magnetic field intensity is lower so the steady state and ELM-like pulse input energy is spread over a larger area, making it easier for the background gas to dissipate it. The energy in the capacitor banks is also lower. ID 5-10 are, conversely, referred to as strong pulses, with stronger magnetic field and higher capacitor bank energy.

Two $\sim40$s long discharges are performed for each condition. In each, first the steady state plasma is generated, then a series of ELM-like pulses are generated on top. More detail on the procedure are given in \autoref{Sampling strategy}.
During weak pulses a small neutral pressure increase was observed from the initial steady state plasma phase to when repeated ELMs are delivered, as it is expected in tokamaks\cite{Smith2020}, but not for strong pulses. This is likely because the lower magnetic field allows the plasma to spread over a larger volume ($\sim$50\% more from OES) and can therefore better fill the target skimmer (5mm diameter), preventing gas to be removed by the heating chamber pump, that always runs at 82\% speed, from the target chamber, whose pump can be run slower to cause a higher target chamber pressures. The pressure increase is from $\sim$1\% in ID4 to $\sim$5\% in ID1. This is likely due to the increase of the gas temperature in the target chamber due to the heating from the ELM-like pulse, as the more gas is present, the lower is the temperature and the less is the pressure increase.

\begin{table*}
\begin{tabular}{ | >{\centering}m{01em} | >{\centering}m{1.5cm}| >{\centering}m{1.5cm} | >{\centering}m{1.2cm} | >{\centering}m{2.2cm} | >{\centering}m{2.2cm} | >{\centering}m{3.6cm} | >{\centering}m{2cm} | } 
  \hline
  ID & capacitor voltage [V] & capacitor energy [J] & Magnetic field [T] & Target chamber $H_2$ feeding [slm] & Target chamber pump speed [\%] & Steady state neutral pressure in target chamber [Pa] & Stage (defined in \autoref{Fast camera}) \tabularnewline 
  \hline
  1 & 370 & 10.3 & 0.6 & 0 & 82 & 0.223 & 1 \tabularnewline 
  \hline
  2 & 370 & 10.3 & 0.6 & 0 & 25 & 0.385 & 1\tabularnewline
  \hline
  3 & 370 & 10.3 & 0.6 & 10 & 25 & 5.991 & 2/3\tabularnewline
  \hline
  4 & 370 & 10.3 & 0.6 & 20 & 25 & 10.956 & 3\tabularnewline
  \hline
  5 & 800 & 48.0 & 1.3 & 0 & 82 & 0.296 & 1\tabularnewline
  \hline
  6 & 800 & 48.0 & 1.3 & 0 & 25 & 0.516 & 1\tabularnewline
  \hline
  7 & 800 & 48.0 & 1.3 & 5 & 25 & 4.370 & 1/2\tabularnewline
  \hline
  8 & 800 & 48.0 & 1.3 & 10 & 25 & 8.170 & 2\tabularnewline
  \hline
  9 & 800 & 48.0 & 1.3 & 15 & 25 & 11.847 & 2\tabularnewline
  \hline
  10 & 800 & 48.0 & 1.3 & 20 & 25 & 15.040 & 2\tabularnewline
  \hline
\end{tabular}
  \caption{Table of the experimental conditions of Magnum-PSI for the experiments presented in this work. Common parameters are: steady state source current 140A to generate the background plasma, 31.2mm target to OES/TS distance,TZM (a molybdenum alloy) target. ID 1-4 are referred to as weak pulse conditions while ID 5-10 are referred to as strong pulses. The stages are 1-3 and correspond to: fully attached (1), partially detached (2) and fully detached (3) as it will be defined in \autoref{Fast camera}.}
  \label{tab:table1}
\end{table*}

\subsection{Diagnostics}\label{Diagnostics}

In this section, the diagnostics employed in this study will be described. Some diagnostics have the time resolution to be able to observe individual ELM-like pulses, while others only a limited part of it, so a sampling strategy was adopted to reconstruct the full behaviour of the ELM-like pulse. More detail is given in \autoref{Sampling strategy}.

The results from the diagnostics used in the experiments are described in the following sections:
\begin{itemize}
    \item[\ref{Fast camera}] Vision Research Phantom v12.1 CMOS fast visible light camera: axial and radial view of the plasma
    \item[\ref{Thomson scattering}] Thomson scattering (TS): electron temperature and density
    \item[\ref{IR camera}] Infrared (IR) camera FLIR SC7500MB: target temperature (used with calibration from FAR-Associate Spectro Pyrometer FMPI)
    \item
     Jarell-Ash, Czerny-Turner optical emission spectrometer (OES): Hydrogen atomic line emission (Balmer series $p=4-8 \rightarrow 2$, used in the second paper too, see \cite{Federici2023c})
    \item
     Power source (ADC): temporal variation of the power delivered to the plasma (used in the second paper too, see \cite{Federici2023c})
\end{itemize}

\subsubsection{Fast Camera}\label{Fast Camera1}
A Vision Research Phantom v12.1 CMOS camera is installed such that it has a radial and axial view of the plasma coming from the source (left) and directed to the target (right). The view is through a lateral viewport of the vacuum vessel so the useful FOV is limited to a diameter of $\sim$75mm (see \autoref{fig:SS}, \ref{fig:ELM1}.) The camera records visible light and it’s not used to deliver quantitative information but it is very useful to establish the qualitative behaviour of the plasma. The frame rate used was 67kHz, enough to resolve individual ELM-like pulses ($\sim$1ms in duration, 67 frames). The information from multiple pulses was then averaged. Thanks to this diagnostic it is possible to identify 3 distinct regimes/stages that are evident as the neutral pressure in the target chamber is increased and therefore increasing the level of detachment.

\subsubsection{Thomson scattering (TS)}\label{Thomson scattering1}
Thomson scattering (TS) is a diagnostic that allows the measurement of $T_e$ and $n_e$ by firing a laser beam through the plasma and collecting the scattered light from a particular measurement volume determine by the intersection of the laser beam with each TS viewing LOS. 50 LOS are measured simultaneously in the radial direction (in the same plane as OES as shown in \autoref{fig:layout}) to reconstruct the profile of the plasma. In Magnum-PSI, TS can be used for steady state plasmas but also for time dependent measurements, albeit with reduced performance. For this campaign the system was used in time dependent mode with $50\mu s$ time resolution and integration time, with an uncertainty $<3\%$ for electron density and $<10\%$ in electron temperature for $n_e>2.8 \cdot 10^{20} \#/m^3$. The time between consecutive measurements must be equal to the laser frequency, 10Hz, so to reconstruct the whole time evolution of the ELM the TS data has to be accumulated over multiple ELM-like pulses with a dedicated sampling strategy, see \autoref{Sampling strategy}. \cite{VanDerMeiden2012}

\subsubsection{IR camera}\label{IR camera1}
This data is collected by a FLIR SC7500MB infrared camera that observes the surface of the target in the wavelength range $3.97-4.01 \mu m$. The camera's view is directed towards the target by a mirror and enters the target chamber from a port upstream from the target as shown in \autoref{fig:layout}. The temperature calibration is performed with a dedicated FAR-Associate Spectro Pyrometer. The FLIR camera operates at a frame rate of about 3kHz, corresponding to $\sim$3 frames per ELM-like pulse, with the integration time set to $100 \mu s$. To increase the confidence and rime resolution of the measurement, records for multiple ELM-like pulses are overlapped and averaged once the target reached a thermal steady state, following a similar procedure to that employed by Li. \cite{Li2020} The IR camera results provide 2D temperature profiles of the target over time. Due to differences in the triggering system, it is not possible to directly correlate IR camera time with OES or TS, so these results are used independently.

\subsubsection{Power supply}\label{Power supply}
The steady state power supply regulates the DC plasma source voltage such that its current is equal to the set point. A capacitor bank composed of 28 individual sections consisting of an inductor (L=160$\mu$H) and a capacitor (C=150$\mu$F) is connected in parallel to the DC source.\cite{Morgan2014} The energy stored in each capacitor (shown in \autoref{tab:table1}) is given by $\frac{1}{2}CV^2$. Each of these capacitors can be individually controlled to charge to a set voltage and discharge at a specified time. The current released by the capacitor will be additional to the steady state one. The voltage and current at the plasma source are recorded and data from multiple ELM-like pulses is overlapped to compensate for small differences in each capacitor/inductor pair in a similar fashion as done for the IR camera. The energy transferred into plasma energy was measured to be 92\% of the electrical energy dissipated at the plasma source during ELM-like pulses.\cite{Morgan2014} Some of the energy can additionally be dissipated in the source and heating chambers volume and on the source and target chamber skimmers, before the plasma enters the target chamber. The pressure in the source and heating chambers is maintained as low as possible via differential pumping to reduce interaction of the beam with cold gas and consequent energy losses. 

\begin{figure*}[!ht]
    \captionsetup{labelfont={color=white}}
     \centering
     \begin{subfigure}{0.35\textwidth}
         \centering
         \vspace*{-0mm}
         \includegraphics[width=\textwidth,trim={26 0 17 11},clip]{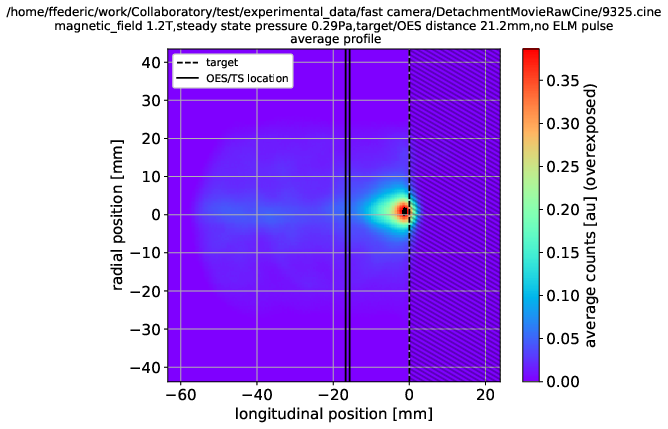}
         \vspace*{-17mm}
         {\color{white}\caption{\phantom{weww}}\label{fig:SSa}}
     \end{subfigure}
     \hfill
     \begin{subfigure}{0.31\textwidth}
         \centering
         \vspace*{-0mm}
         \includegraphics[width=\textwidth,trim={39 0 17 11},clip]{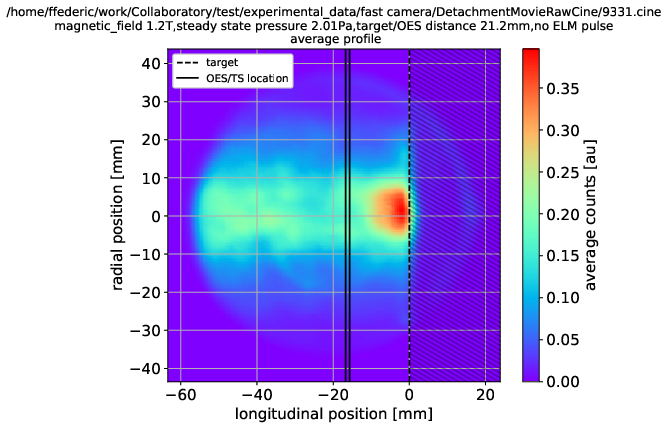}
         \vspace*{-17mm}
         {\color{white}\caption{\phantom{wewwwww}}\label{fig:SSb}}
     \end{subfigure}
     \hfill
     \begin{subfigure}{0.31\textwidth}
         \centering
         \vspace*{-0mm}
         \includegraphics[width=\textwidth,trim={39 0 17 11},clip]{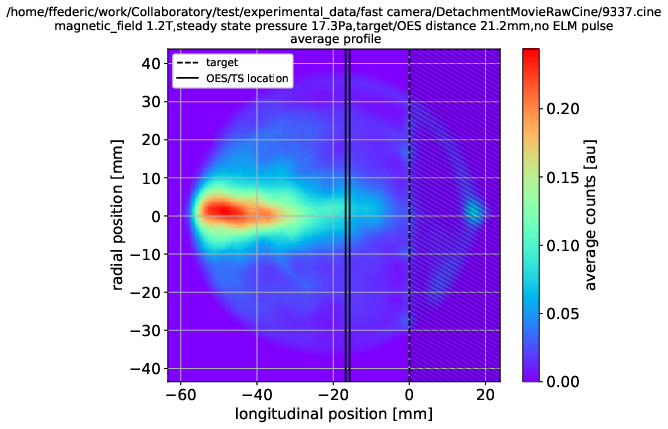}
         \vspace*{-17mm}
         {\color{white}\caption{\phantom{wewwwww}}\label{fig:SSc}}
     \end{subfigure}
        \captionsetup{labelfont={color=black}}
        \vspace*{+6mm}
        \caption{Visible fast camera brightness images corresponding to a steady state plasma in conditions similar to this work from an experiment by Akkermans\cite{Akkermans2020} (1.2T, 120A, H$\alpha$ filter present, (\subref{fig:SSa}) is affected by overexposure in the black region near the target). The target chamber neutral pressure is increased to induce detachment and corresponds to 0.29Pa (\subref{fig:SSa}), 2.01Pa (\subref{fig:SSb}) and 17.3Pa (\subref{fig:SSc}). The shaded region indicates the position of the target. The OES and TS measurement locations in z are indicated by the 2 vertical lines. The fast camera view is through a lateral viewport, giving a useful FOV of $\sim$75mm diameter.}
        \label{fig:SS}
\end{figure*}

The results that can be gathered by the examination of the data from individual diagnostics will now be presented. The fast camera will be used to identify the spatial features of the radiation profile during each pulse, which will be used to characterize the level of detachment and to distinguish the influence of increasing neutral pressure on the burn through of the ELM-like pulse. Thomson scattering allows to measure the main plasma properties so that the plasma pressure losses can be compared with target chamber neutral pressure. The target surface temperature measured with the IR camera allows to determine the energy delivered to the target by the ELM-like pulse, finding it decreasing with increasing neutral pressure until full dissipation of the pulse energy is achieved.\cite{Federici2023d}

The understanding will be deepened in the second publication related to this study with the use of a Bayesian technique specifically developed to interpret in a self-consistent way results from different diagnostics. This allows to quantify the relevance of molecular assisted processes compared to atomic ones.\cite{Federici2023c}

\section{Fast camera}\label{Fast camera}
In this section the visible fast camera brightness is used to examine the spatial distribution of the emission in the target chamber and how the ELM-like progress towards the target is impacted by increasing neutral pressure. This will allow us to define 3 progressive stages of detachment based on the averaged spatial brightness profile.

\subsection{Steady state}\label{Steady state}

In order to characterise the state of the plasma during the ELM-like pulse from the visible fast camera brightness, the known characteristics of a steady state plasma are shown in \autoref{fig:SS}. \cite{Akkermans2020,Perillo2019}
These images are taken with a H$\alpha$ filter, with the plasma in similar conditions to the ones in this work (1.2T, 120A) during an experiment by Akkermans.\cite{Akkermans2020} Because of the presence of the filter the images are blurred.

For low target chamber neutral pressure, the plasma emitting region is concentrated close to the target at the centre of the plasma column (\autoref{fig:SSa}). This indicates that the interactions between plasma and the background gas are weak and the plasma can flow towards the target mostly unperturbed. This is due both to the low neutral density and to the fact that the high plasma temperature typical of low target chamber neutral pressure causes further neutral hydrogen dilution. \cite{DenHarder2015} 

For increased neutral pressure (\autoref{fig:SSb}) the radiation in the volume increases while it decreases close to the target; the plasma volume away from the target plays an increasingly important role. By increasing the neutral pressure even further (\autoref{fig:SSc}) the radiation near the target is strongly reduced and the bulk of the plasma radiation recedes from the target. The peak plasma density close to the target measured by TS first increases from 1.13 to $1.98 \cdot 10^{20}\#/m^3$ from 0.29 to 2.01Pa, and then decreases to the point that TS measurement fails at 17.3Pa. The peak plasma temperature conversely decreases from 3.34eV to 1.53eV and becomes undetectable at 17.3Pa. The characteristics of the plasma near the target from TS thus display a monotonic loss of plasma pressure as neutral pressure increases, consistent with the behaviour of a plasma that experiences detachment caused by the increase of neutral pressure.\cite{Perillo2019} The plasma is classified as attached to the target in \autoref{fig:SSa} and \subref{fig:SSb}, while detached in \autoref{fig:SSc}.

\begin{figure}
	\centering
	\includegraphics[width=\linewidth,trim={7 0 9 45},clip]{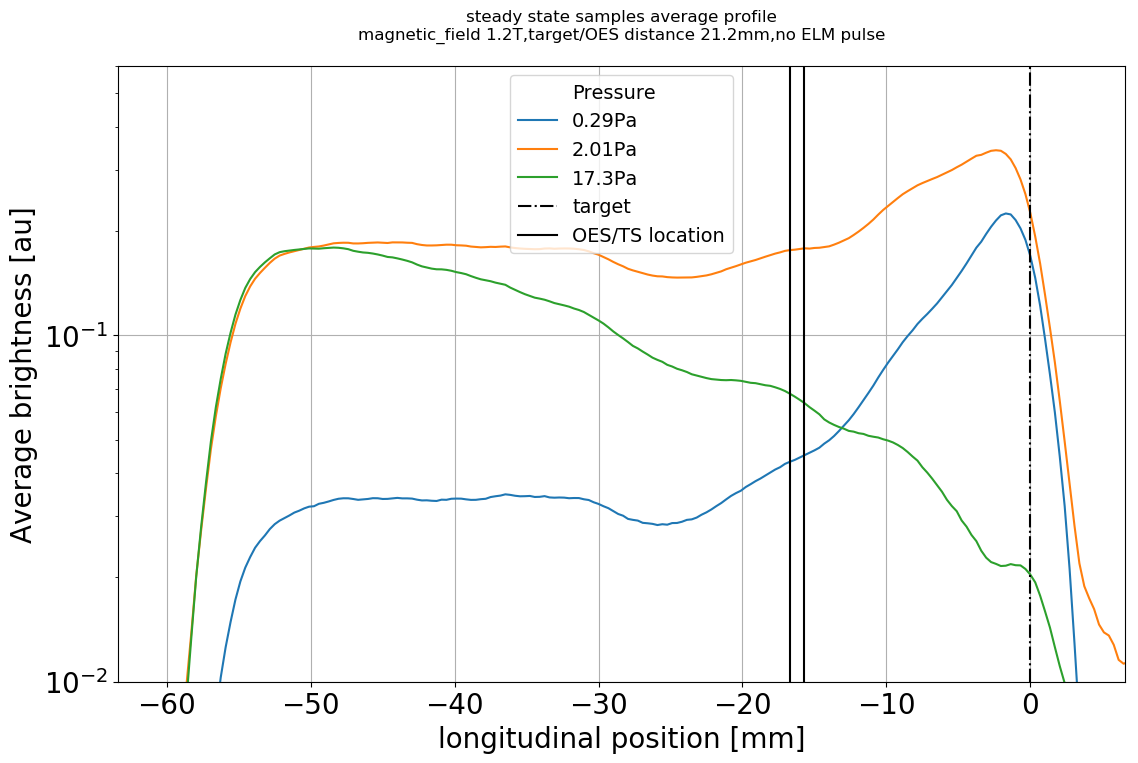}
	\caption{Radial average of the visible fast camera brightness for the steady state cases in \autoref{fig:SS}. The longitudinal position axis is cut after the target. The brightness decreases towards the source as the plasma exits the field of view of the camera.}
	\label{fig:SS2}
\end{figure}

\begin{figure*}
    \captionsetup{labelfont={color=white}}
     \centering
     \begin{subfigure}{0.36\textwidth}
         \centering
         \vspace*{-0mm}
         \includegraphics[width=\textwidth,trim={20 0 8 8},clip]{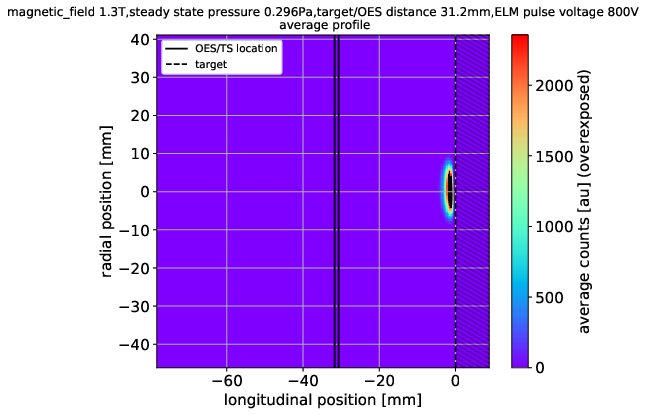}
         \vspace*{-17mm}
         {\color{white}\caption{\phantom{weww}}\label{fig:ELMa}}
     \end{subfigure}
     \hfill
     \begin{subfigure}{0.31\textwidth}
         \centering
         \vspace*{-0mm}
         \includegraphics[width=\textwidth,trim={34 0 8 7.5},clip]{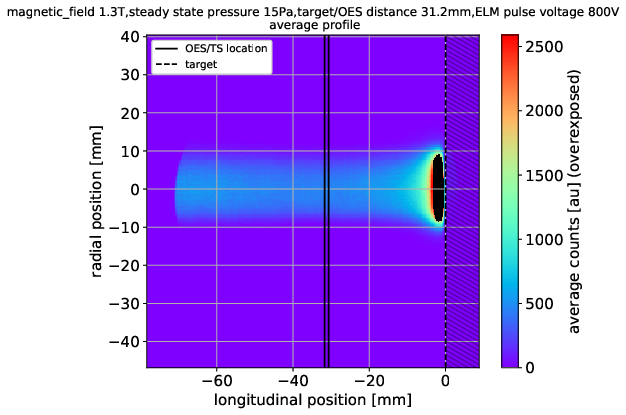}
         \vspace*{-17mm}
         {\color{white}\caption{\phantom{wewwwww}}\label{fig:ELMb}}
     \end{subfigure}
     \hfill
     \begin{subfigure}{0.31\textwidth}
         \centering
         \vspace*{-0mm}
         \includegraphics[width=\textwidth,trim={34 0 8 8},clip]{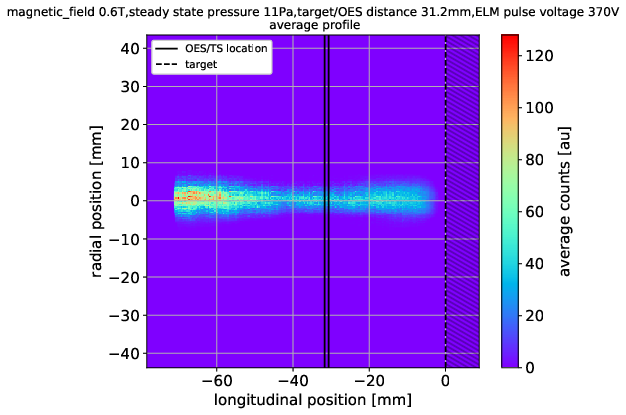}
         \vspace*{-17mm}
         {\color{white}\caption{\phantom{wewwwww}}\label{fig:ELMc}}
     \end{subfigure}
        \captionsetup{labelfont={color=black}}
        \vspace*{+6mm}
        \caption{Fast visible camera images of the ELM-like obtained by averaging temporally over the pulse duration. The images show a plasma which is increasingly detached and corresponds to ID 5 (\subref{fig:ELMa}), 10 (\subref{fig:ELMb}), 4 (\subref{fig:ELMc}) in \autoref{tab:table1} ((\subref{fig:ELMa}) and (\subref{fig:ELMb}) are affected by overexposure in the black region near the target). The fast camera view is through a lateral viewport, giving a useful FOV of $\sim$75mm diameter.}
        \label{fig:ELM1}
\end{figure*}

By averaging the visible fast camera brightness over the radial direction one obtains the profiles in \autoref{fig:SS2}. Those profiles reinforce how the main interactions of the plasma shift from close to the target to the volume of the plasma column for increasing neutral pressure. The increase in visible radiation correlates well with observations from simulations that H$\alpha$ emission is greatly increased from the detachment onset onward because of the influence of molecular reactions.\cite{Zhou2022} Using this method of determining the detachment state of the plasma, we can proceed interpreting the visible fast camera brightnesses during the ELM-like pulses.

\subsection{Effect of ELM-like pulses}\label{Effect of ELM-like pulses}

For this study, the fast camera was set with a high frame rate of 67kHz and low integration time of 900ns in order to capture the bright ELM-like pulse behaviour. With those settings, the brightness of the steady state is not detectable and only the ELM-like pulses can generate counts. The ELM-like pulse lasts 0.7 to 1ms (equivalent to 67-47 frames) depending on the target chamber neutral pressure and a clear movement of a radiation front from the source to the target could not be clearly identified with this settings. In order to extract information on the behaviour across the ELM-like pulse the camera measurement is averaged in time over the time interval with non zero data. The data from a few ELM-like pulses in the same conditions is then further averaged. A sample of the results for the plasma conditions in this paper are shown in figure \autoref{fig:ELM1}.

\begin{figure}
	\centering
	\includegraphics[width=\linewidth,trim={110 0 170 85},clip]{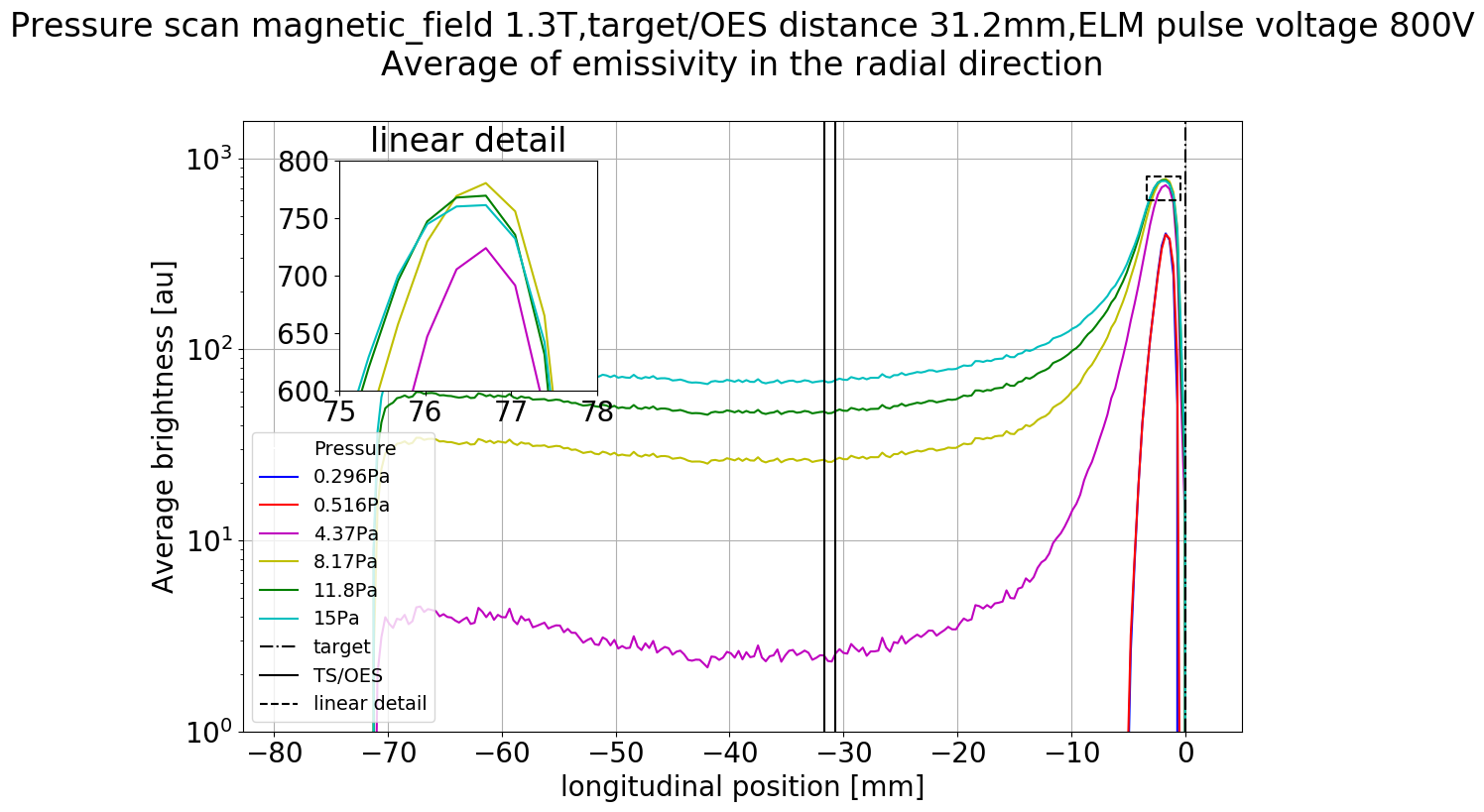}
	\caption{Radial and temporal average of the visible fast camera brightness for the neutral pressure scan with strong ELM-like pulses, where we observe the transition from Stage 1 to 2. (ID 5-10 in \autoref{tab:table1}).}
	\label{fig:ELM2}
\end{figure}

\autoref{fig:ELMa} corresponds to a low target chamber neutral pressure. Under such conditions, it is typical to observe a bright emission in proximity of the target while almost nothing is visible far from it; the vast majority of the plasma-neutral interactions and the resulting emission happens at the target while the plasma is transported, mostly undisturbed, through the bulk of the target chamber. This closely resembles what was observed in \autoref{fig:SSa}, indicating that the plasma is, in this case, attached to the target. Another observation is that, during the ELM-like pulse, the visible fast camera brightness close to the target increases and then decreases, while maintaining the same relative shape (see \autoref{fig:ELM2}). This points towards the fact that the plasma is attached before, during and after the pulse. This is also confirmed by independent TS and OES measurements away from the target. From the steady state to the peak ELM-like pulse values $T_e$ rises from $\sim$5 to 9.4eV and $n_e$ from $\sim$1 to $23 \cdot10^{20}\#/m^3$ (see \autoref{Thomson scattering} for further details). This experimental condition is the first one encountered starting from low neutral pressure, so it is referred to a Stage 1. 

\begin{figure}
	\centering
	\includegraphics[width=\linewidth,trim={110 0 170 85},clip]{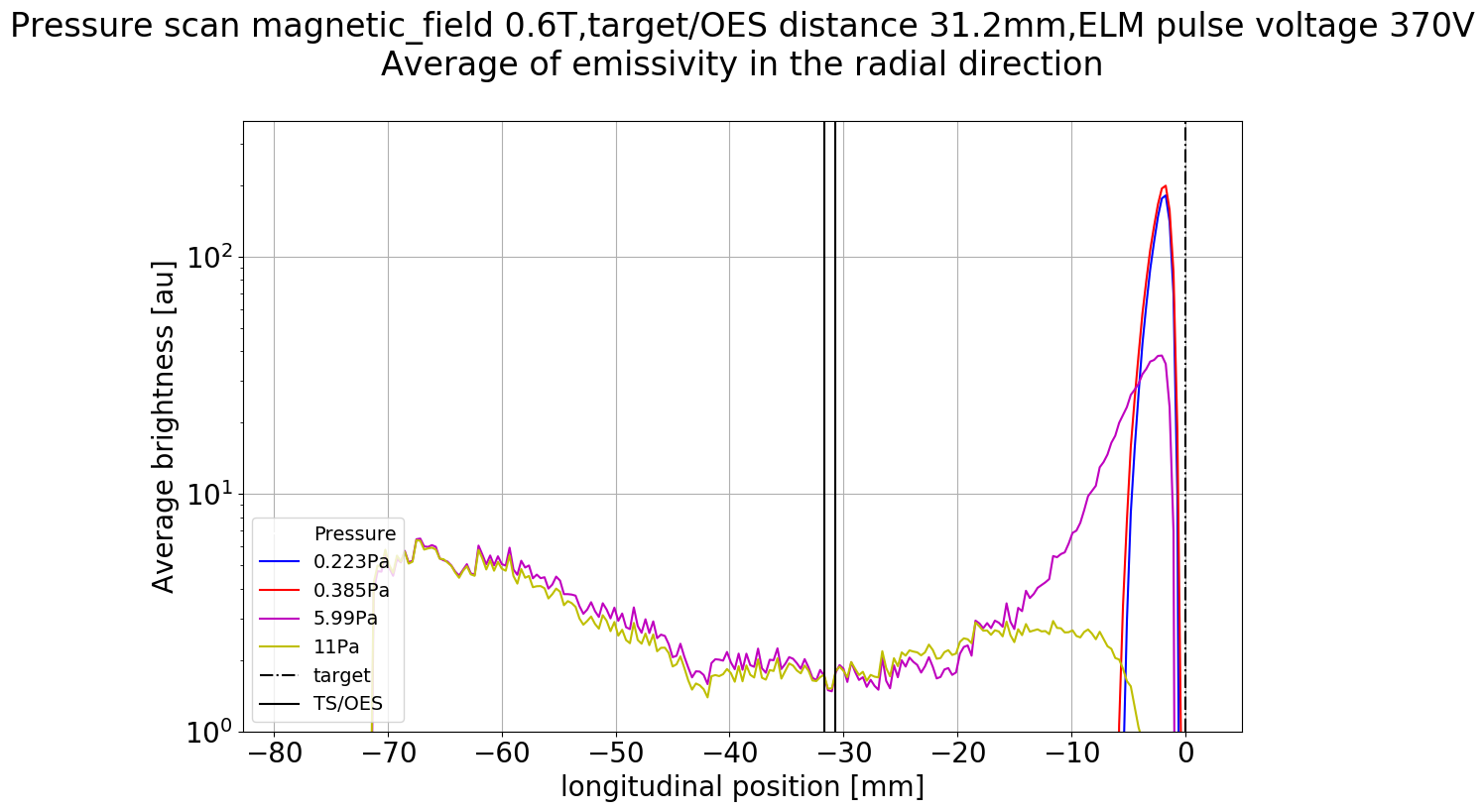}
	\caption{Radial and temporal average of the visible fast camera brightness for the neutral pressure scan with weak ELM-like pulses, where we observe the transition from Stage 1 to 3 (ID 1-4 in \autoref{tab:table1}).}
	\label{fig:ELM3}
\end{figure}

In \autoref{fig:ELMb} the neutral density is increased. The light emission in the bulk of the target chamber increases with respect to at the target and is typically quite homogeneous along the magnetic field direction. The emission closely resembles \autoref{fig:SSb}, meaning that during the ELM-like pulse the plasma is still attached to the target. This in not the case before and after, where the density is so low that TS measurements fail and no emission is visible from OES. During the ELM-Like pulse, the peak $T_e$ decreases to $\sim 4.6$eV while $n_e$ increases to $\sim 52 \cdot 10^{20}\#/m^3$. This condition is referred to as Stage 2.

If the neutral pressure is further increased from what causes Stage 2 to occur, the dissipation of the plasma in the volume, also referred as baffling, in the target chamber can be so severe that the strong luminous spot close to the target might not appear. This is shown in \autoref{fig:ELMc}. In such cases the interaction of the plasma in the volume is so strong as to dissipate the ELM-like pulse almost entirely before it reaches the target. Only luminosity in the volume is present, closely resembling \autoref{fig:SSc} which corresponds to a detached plasma, meaning the plasma is detached before and during the ELM-like pulse. The OES, TS, and visible fast camera brightness is measurable for a much shorter time compared to the ELM-like pulse duration ($\sim 300 \mu s$ compared to a pulse length of $700 \mu s$), meaning only the most intense part of the ELM-like pulse makes it to the measuring location and the energy is dissipated in the volume. This is further confirmed by TS as peak $T_e$ drops from 0.7 to 0.4eV and $n_e$ from 14 to $5 \cdot10^{20}\#/m^3$ increasing the neutral pressure from Stage 2 (ID 3 to 4 in \autoref{tab:table1}). This regime is referred to as Stage 3.

The information in \autoref{fig:ELM1} is summarised in two figures: \autoref{fig:ELM2} and \autoref{fig:ELM3} (averaged radially and in time for each pressure) which displays the average of the visible fast camera brightness as a function distance from the target. Both figures include the analysis for several neutral pressures for the two scans included in this work (ID 1 to 4 and 5 to 10 in \autoref{tab:table1}).

\autoref{fig:ELM2} corresponds to a strong ELM-like pulse. The emission is always peaked at the target while the emission away from it increases significantly for increasing pressure. At the two lowest neutral pressure settings the plasma is in Stage 1, it is at an intermediate state for ID7 (with a significantly lower emission in the volume then the other higher pressure cases) while it then transitions to Stage 2 for ID8-10 in \autoref{tab:table1}. The brightness becomes more homogeneous along the magnetic field for increasing neutral pressure, reducing the signal anisotropy in the longitudinal direction. 

The emission at the peak in \autoref{fig:ELM2} initially increases with neutral pressure but, as shown in the linear detail it then starts to decrease, signalling the start of the transition from Stage 2 to 3 (it must be noted, though, that the peak region is affected by saturation). It will be shown in \autoref{Thomson scattering} that the peak plasma pressure measured by TS does not significantly change in the transition from Stage 1 to 2, indicating that the plasma is still attached during all the strong ELM-like pulses.

\autoref{fig:ELM3} displays the average of the brightness for the neutral pressure scan with weak ELM-like pulses (ID 1-4 in \autoref{tab:table1}), meaning the pulses are weak compared to the baffling of the background gas. Here we have the transition from Stage 1 (0.223 and 0.385Pa) to 2/3 (5.99Pa) and ultimately 3 (11Pa). The same trend in \autoref{fig:ELM2} continues here, with an even more uniform light emission at the highest pressures. Note that this dataset is unaffected by saturation. The 5.99Pa case is identified as Stage 2/3 because even if some emission comes from nearby the target, that is significantly reduced compared to the Stage 1 cases, and looking at the temporal evolution of the brightness, the pulse reattaches to the target only briefly. Later analysis will show that with neutral pressure 5.99Pa no heat is delivered to the target, typical of Stage 3 rather than 2, and the plasma pressure loss is significant, indicating increasing levels of detachment.

\begin{figure*}[ht!]
    \captionsetup{labelfont={color=white}}
     \centering
     \begin{subfigure}{0.34\textwidth}
         \centering
         \vspace*{-0mm}
         \includegraphics[width=\textwidth,trim={60 0 134 25},clip]{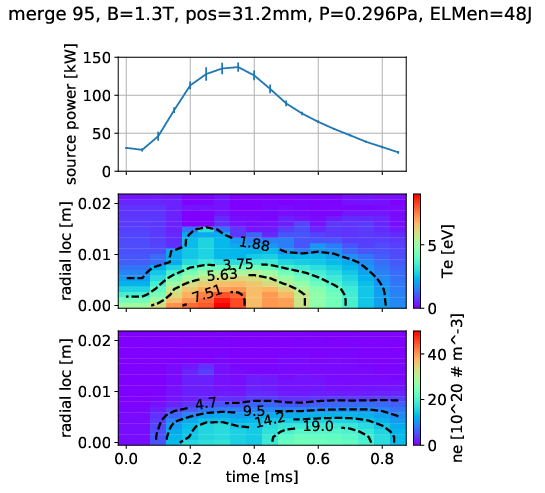}
         \vspace*{-55mm}
         {\color{white}\caption{\phantom{ }}\label{fig:TSb}}
     \end{subfigure}
     \hfill
     \begin{subfigure}{0.285\textwidth}
         \centering
         \vspace*{-0mm}
         \includegraphics[width=\textwidth,trim={100 0 120 25},clip]{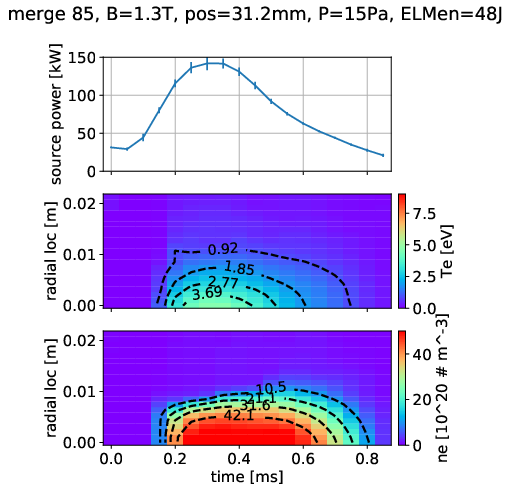}
         \vspace*{-55mm}
         {\color{white}\caption{\phantom{ }}\label{fig:TSa}}
     \end{subfigure}
     \hfill
     \begin{subfigure}{0.341\textwidth}
         \centering
         \vspace*{-0mm}
         \includegraphics[width=\textwidth,trim={110 0 70 25},clip]{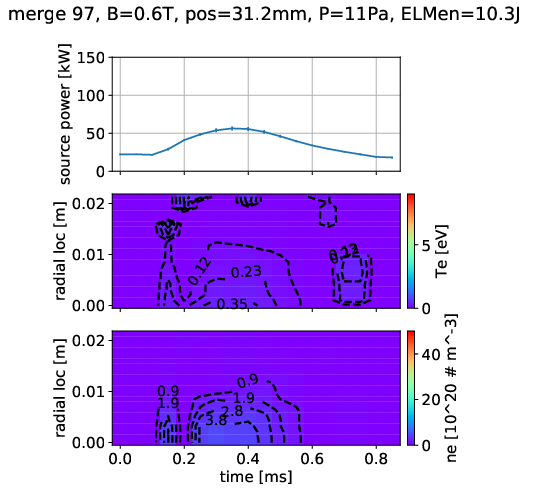}
         \vspace*{-55mm}
         {\color{white}\caption{\phantom{ }}\label{fig:TSc}}
     \end{subfigure}
        \captionsetup{labelfont={color=black}}
        \vspace*{+45mm}
        \caption{Typical source power (top), Thomson scattering $T_e$ (mid) and $n_e$ (bottom) measurement for strong pulses in Stage 1 (\subref{fig:TSb}), Stage 2 (\subref{fig:TSa}) and weak pulses in Stage 3 (\subref{fig:TSc}) (ID 5, 10, 4 in \autoref{tab:table1} respectively). TS data is smoothed over time and radius in order to match the same steps of OES.
        }
        \label{fig:TS1}
\end{figure*}

The distinction in stages helps to discriminate between types of behaviour of the plasma and its influence on other diagnostics and the power balance: 
\begin{enumerate}
    \item[Stage 1] The plasma is attached to the target before, during and after the ELM-like pulse. These conditions are typically correlated to low volumetric losses, high steady state target temperature and measurable steady state plasma conditions, as it will be clear from analysis of the IR camera and TS data. 
    \item[Stage 2] The plasma is detached from the target, but it reattaches during the ELM-like pulse. The target is cold in between ELM-like pulses but a significant target heating can be provided transiently and volumetric losses dominate as a power loss mechanism.
    \item[Stage 3] The plasma is alwaus detached from the target, also during the ELM-like pulse. The lack of emission near the target coincides with negligible target heating, significant plasma losses before reaching the OES/TS location and an increased anisotropy in the plasma column.
\end{enumerate}

As previously mentioned the emission is not homogeneous in the whole target chamber. Strong emission is located close to the target and even in the bulk there is a slight increase in emission moving away from the target. The anisotropy in the bulk of the target chamber, though, is weak and decreasing for increasing neutral pressure. \emph{This observation will be used in the second paper part of this study\cite{Federici2023c} to support the approximation that volumetric power losses can be considered constant from skimmer to target.}

\section{Thomson scattering}\label{Thomson scattering}

The Thomson scattering diagnostic measures the main plasma properties ($T_e$, $n_e$) and allows to determine the temporal evolution of the ELM-like pulse and the transfer of plasma energy from high temperature (and low density) to high density (and low temperature) for increasing neutral pressure. The LOSs of the light collection system are centered on the plasma column so data is collected on the plasma above and below the plasma column centre. These are averaged to return the radial profiles. A small time difference between OES and TS is also present so the raw TS $T_e$ and $n_e$ are resampled and smoothed, to match the OES time and spatial bases and obtain the temporal and radial profiles.

In \autoref{fig:TS1} are shown typical $T_e$ and $n_e$ from TS for experimental conditions in Stage 1, 2 and 3. The upper row shows the power dissipated by the plasma source averaged over multiple ELM-like pulses and aligned with TS. As mentioned before, the weak pulses are obtained via a lower magnetic field and a lower energy of the pulse itself. The lower magnetic field causes the energy of the pulse to distribute over a larger area (radius) and increase the volume of contact between plasma and cold gas. The lower temperature and density also causes the neutral mean free path to increase, allowing them to penetrate deeper inside the plasma column. For strong pulses $T_e$ decreases while $n_e$ increases as the neutral pressure and the amount of gas available to be ionised increases. The steady state properties become undetectable while the time corresponding to peak density shifts earlier in time. For weak pulses and high neutral pressure peak density is at even earlier times with a very low plasma temperature.

\begin{figure}
	\centering
	\includegraphics[width=\linewidth,trim={0 0 0 0},clip]{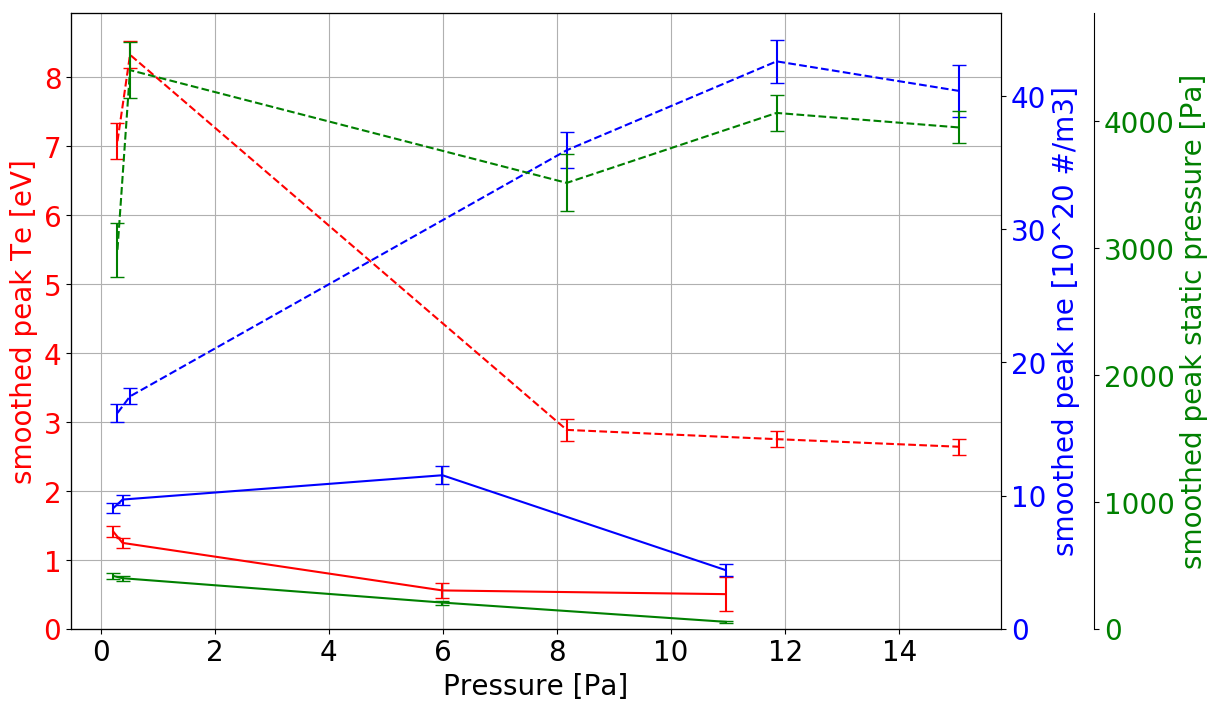}
	\caption{Comparison of plasma peak $T_e$, $n_e$ and $p_e$ smoothed over $700 \mu s$ for a neutral pressure scan with strong pulses (dashed lines, B=1.3T, pulse energy=48J), corresponding to \autoref{fig:ELM2} and weak (solid lines, B=0.6T, pulse energy=10.3J), corresponding to \autoref{fig:ELM3} (ID 5-10 and 1-4 respectively in \autoref{tab:table1}).}
	\label{fig:TS2}
\end{figure}

A comparison of the peak $T_e$, $n_e$ and $p_e$ ($=n_e2T_e k_B$) smoothed over $700 \mu s$ (the duration of the ELM pulse) is shown in \autoref{fig:TS2}. For strong pulses (dashed lines) $T_e$ decreases while $n_e$ increases as the neutral pressure and the amount of gas available to be ionised increases. Only the maximum neutral pressure achievable is high enough to cause a possible decrease of the measured $n_e$, indicating that the plasma may be starting to transition to a detached state also during the strong ELM-like pulses (as suggested by the fast camera results). This trend can be seen also in the static plasma pressure, that remains at similar levels as the neutral pressure increases. If it could have been further increased the plasma pressure would likely have decreased as would both $n_e$ and $T_e$. 

Similar trends can be seen for weak pulses too (solid lines), but $n_e$ starts decreasing at much lower neutral pressure, indicating that it is easier for the gas to dissipate the weak ELM plasma energy. In the highest pressure case the plasma is significantly depleted in the volume before reaching the TS location, causing the TS measurement itself to fail except at the very peak of the ELM-like pulse, and not be representative of the whole plasma column. For this reason weak pulses are not suited to be analysed with the Bayesian model in \cite{Federici2023c}. This situation corresponds to an ELM-like pulse in Stage 3 as defined before.

In summary increasing the neutral pressure (ID 5-10 in \autoref{tab:table1}) the plasma transitions from Stage 1 to Stage 2: the energy losses before reaching the TS measuring location increase but the ELM-like pulse energy is still enough to maintain the plasma pressure constant. Further increasing the neutral pressure (ID 1-4 in \autoref{tab:table1}) the energy losses for the weak ELM pulses keep increasing and $n_e$ as well as $p_e$ decrease transitioning to a detached plasma in Stage 3.



\section{IR camera}\label{IR camera}
The IR camera returns 2D profiles of the target temperature over time. Only 3 frames correspond to a single ELM-like pulse so multiple pulses are overlapped and averaged to increase the resolution aligning all pulses peak as done by Li.\cite{Li2020} To avoid being affected by thermal transients, only the last 150 of the 300 pulses within every scan is considered (see \autoref{Target temperature profile interpretation} for details). From this data, the time dependent peak temperature and the shape of the affected region are extracted. These are used to show that the ELM-like pulse energy delivered to the target decreases with increasing neutral pressure to the point of being negligible therefore achieving the baffling of the ELM-like pulse.

\begin{figure}
     \centering
     \hspace{+11mm}
     \begin{subfigure}{0.705\linewidth}
         \centering
         \includegraphics[width=\textwidth,trim={6 322 535 65.5},clip]{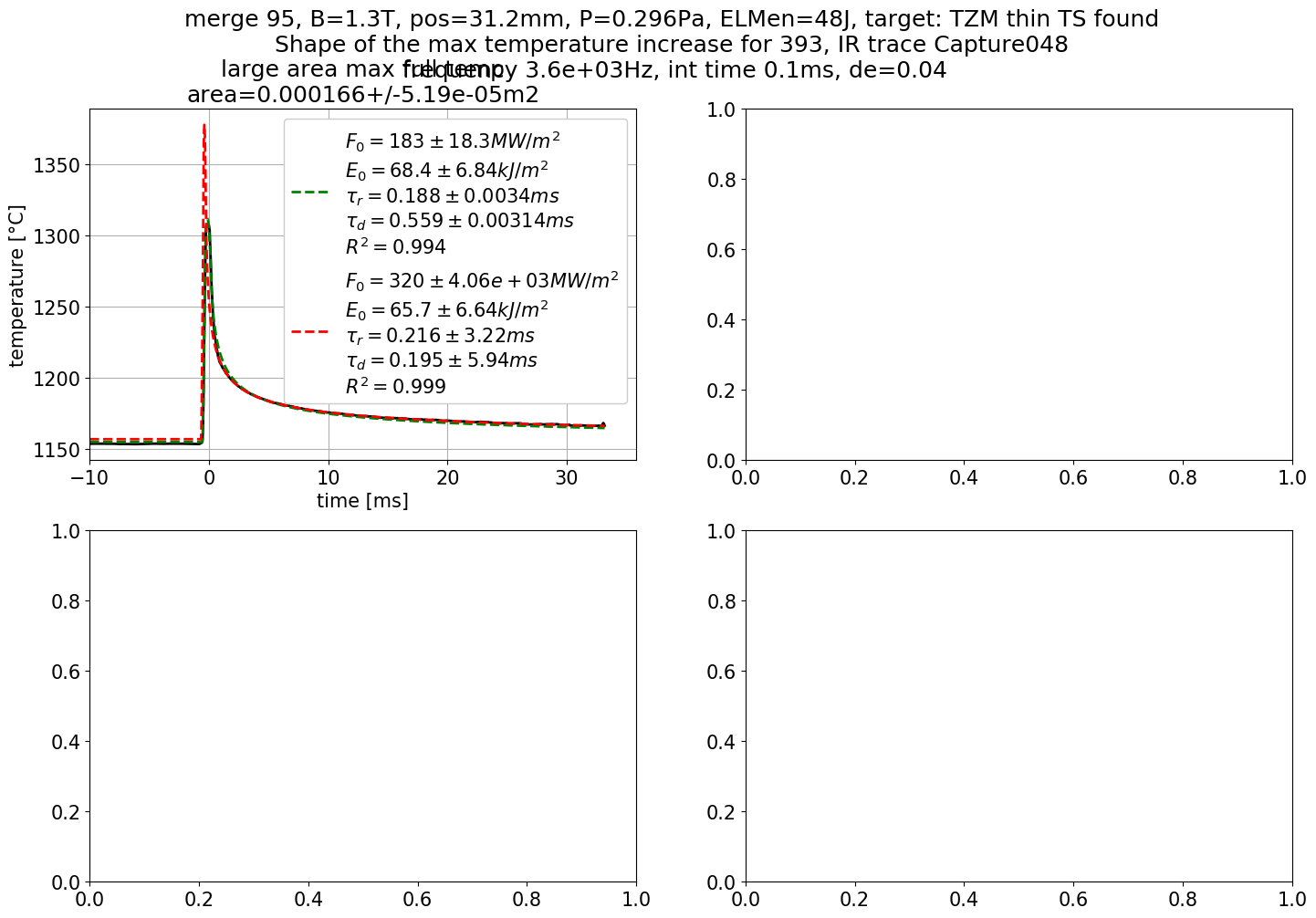}
         \vspace{-20mm}
         \caption{\phantom{wewwwwwwwwwww}}
         \label{fig:IR1a}
     \end{subfigure}
     \hfill
     \vspace*{+12mm}
     \hfill
     \centering
     \begin{subfigure}{0.7\linewidth}
         \centering
         \includegraphics[width=\textwidth,trim={6 322 535 65.5},clip]{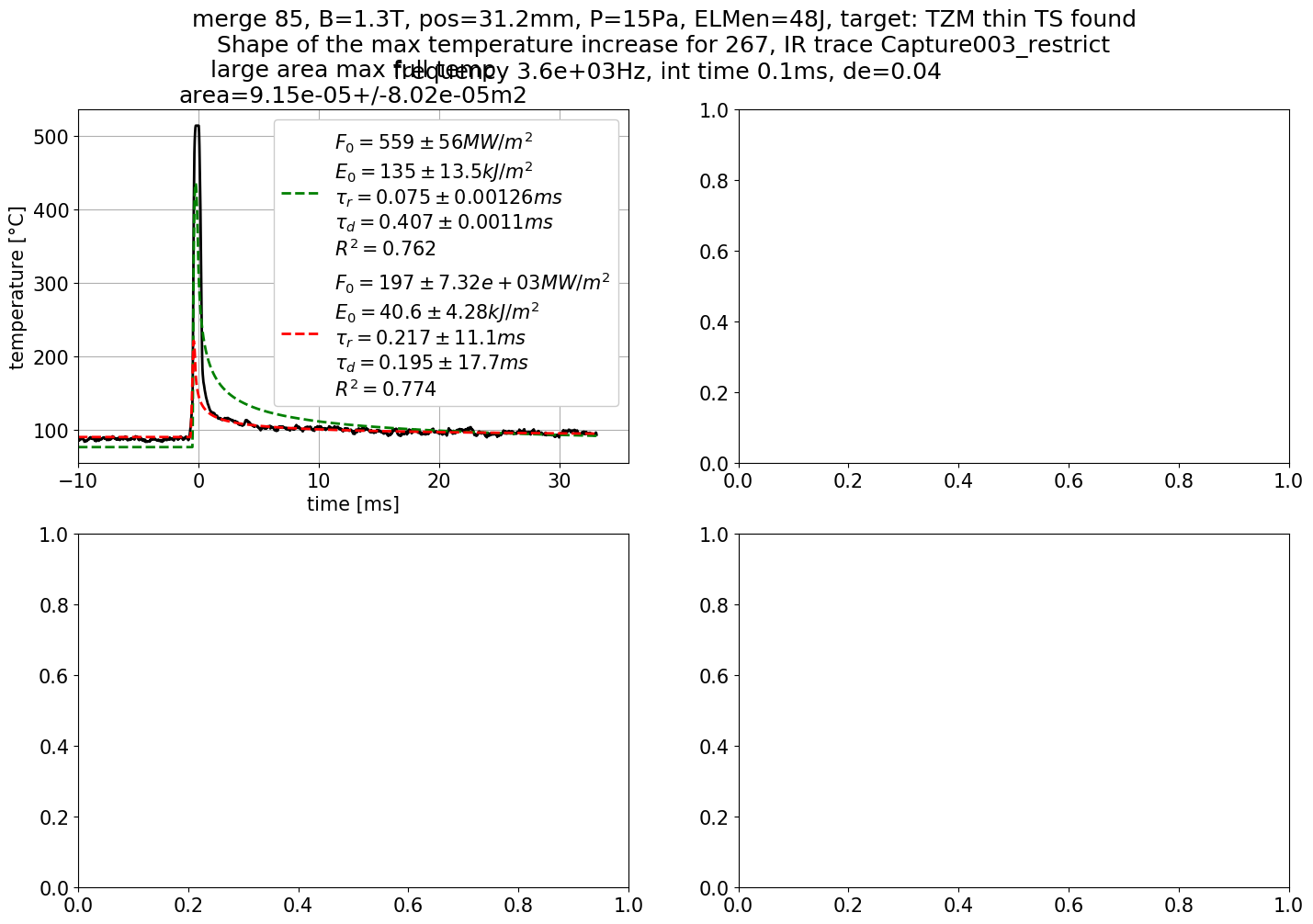}
         \vspace{-20mm}
         \caption{\phantom{wewwwwwwwwwww}}
         \label{fig:IR1b}
     \end{subfigure}
        \vspace{+10mm}
        \caption{Comparison of the measured and fitted time dependence of the peak target temperature: (\subref{fig:IR1a}) Stage 1 case, (\subref{fig:IR1b}) Stage 2 case (ID 5 and 10 in \autoref{tab:table1} respectively). The black curve is the target peak temperature The green dashed line (color online only) is the temperature fit using the entire temperature curve. The red dashed line corresponds to fitting the temperature only for the time 1.5ms after the temperature peak. The curve is then plotted over the entire time domain using the fit parameters. In both cases the analytical model used is Model3 (\autoref{eq:triang} in \autoref{Target temperature profile interpretation}). The fit parameters are shown in the figure legend.}
        \label{fig:IR1}
\end{figure}

To estimate the energy delivered to the target the decrease of the peak target temperature after the pulse is fitted with an analytical solution of the heat transfer equation. Various analytic solutions are available with different degrees of sophistication of the external heat source time and spatial profile. It is important to note that the quantity of interest is an estimation of the total energy density [$J/m^2$] of the heat absorbed by the target from the ELM-like pulse. The analytical solutions considered have the following power density profiles in time and space:

\begin{itemize}
    \item[Model1] a temporally square wave delivered uniformly on a semi infinite plane as described by Behrisch \cite{Behrisch1980}, commonly used in tokamaks to asses the impact of ELMs because of the large extension of the wetted area;
    \item[Model2] a temporally square wave delivered by a spatially Gaussian beam using a solution from Bäuerle \cite{Bauerle2011,Yu2016};
    \item[Model3] a temporally triangular wave delivered by a spatially Gaussian beam using methods from Bäuerle, that better reproduce the Magnum-PSI conditions.
\end{itemize}
All the above analysis methods return similar values for the total energy density delivered to the target by each ELM pulse (details of the comparison of the various techniques are given in \autoref{Target temperature profile interpretation}). The total energy delivered to the target by the ELM-like pulse is then obtained by multiplying the peak energy density ($E_0$) by the affected area found by fitting the temperature profile after the pulse with a Gaussian profile. In \autoref{Target temperature profile interpretation} the MSC.Marc/Mentat® non linear FEM suite was used to reproduce a heat pulse as close as possible to the typical experimental conditions in terms of temporal and spatial variation and temperature dependent material properties. It was found that fitting the temperature after 1.5ms from the peak temperature returns a value of energy delivered within $\pm10\%$ of the real one. It was also found that an uncertainty of $\mp1.5ms$ in the time corresponding to the temperature peak translates to a $\pm10\%$ variation of the energy detected. This translates to a total uncertainty from the method of about $\pm20\%$, good enough for our purposes.

Typical time dependent peak temperatures are shown in \autoref{fig:IR1}. The traces presents a sharp peak followed by a slow cooling period. This fast peak can be caused by emission from the plasma (prompt emission) or an actual temperature increase of the target. For low target chamber neutral pressure the peak temperature matches well with the analytical model: sudden heating of a very thin layer of the surface of the target then the heat quickly redistributes over a larger thickness and radius, being then more slowly dissipated to the actively cooled back plate.\cite{Li2020,Morgan2020} 

In high neutral pressure conditions the peak is still present but it is inconsistent with the following slower cooling. The profiles are fit with Model3, fitting for all times or only the times beyond 1.5ms after the peak. For the Stage 1 case both fits return similar energy values and, most importantly, both fit well the curve for t$>$1.5ms. This is not the case in Stage 2, where the fit for all time fails appreciably. This demonstrates the inconsistency between the temperature peak and the slowly decreasing curve in high neutral pressure conditions. We attribute the enhanced peak to prompt IR line or continuum emission from the plasma (some hydrogen molecular lines lie inside the band allowed by the IR camera filter\cite{Sternberg1989}). This is further reinforced by examining the spatial temperature distribution at the peak compared to afterwards. Further details on this and other aspects of the estimation of the energy delivered to the target are given in \autoref{Target temperature profile interpretation}. The OES could be used to differentiate between emission from the plasma column and then reflected by the target or emission from its immediate vicinity but because of the different wavelength ($3.97-4.01\mu m$ vs $0.32-0.52\mu m$) a direct comparison is not possible. In order to avoid possible misinterpretations only the temperature after 1.5ms from the peak is used rather than the peak itself. This phenomena is also observed in tokamaks, where IR measurements are routinely found unreliable during strong detached phases\cite{Fevrier2020}, and is sometimes used as a marker for detachment itself, or after ELM crashes\cite{Nille2020}.

The plasma steady state heating contribution is estimated by averaging the temperature profile of the target before the ELM-like pulses. Heat conduction through the thin target to the cooled back plate is then calculated, integrated over the target surface, multiplied by the ELM-like pulse duration, and added to the time dependent component to calculate the total heat transferred to the target.

\begin{figure}
	\centering
	\includegraphics[width=\linewidth,trim={35 0 75 22},clip]{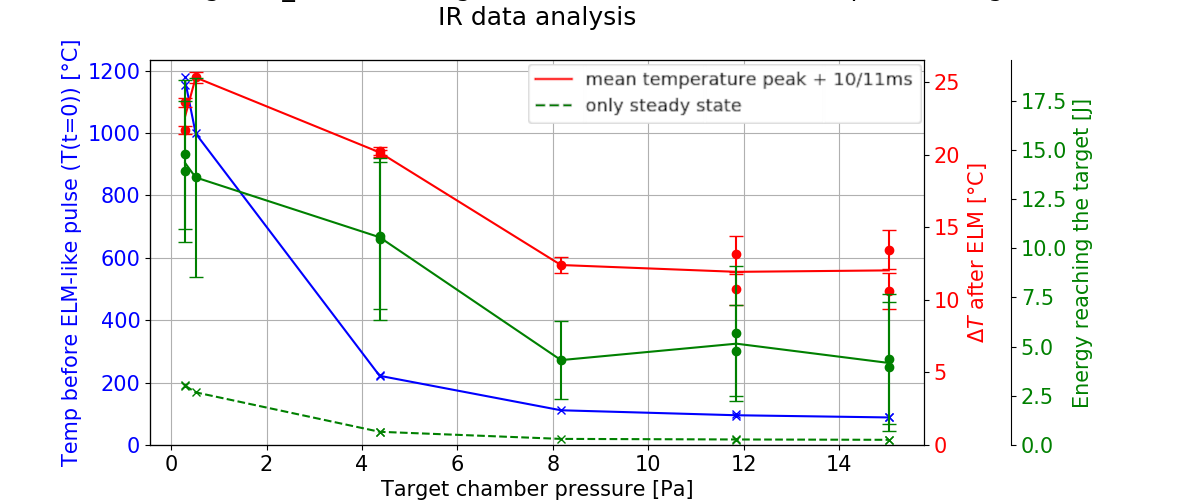}
	\caption{Results of analysis of the IR data for the variations in target chamber neutral pressure. In this figure we focus on the case of strong pulses (48J at the source) and higher magnetic field that yields higher plasma pressure background plasma (ID 5 to 10 in \autoref{tab:table1}). Analyzed data displayed are: target temperature before the ELM (blue), temperature increase during the ELM (red), ELM energy reaching the target (solid green) estimated with Model3 fitted after 1.5ms of the temperature peak and the steady state heat flux component (dashed green).}
	\label{fig:IR2}
\end{figure}
\begin{figure}
	\centering
	\includegraphics[width=\linewidth,trim={35 0 80 30},clip]{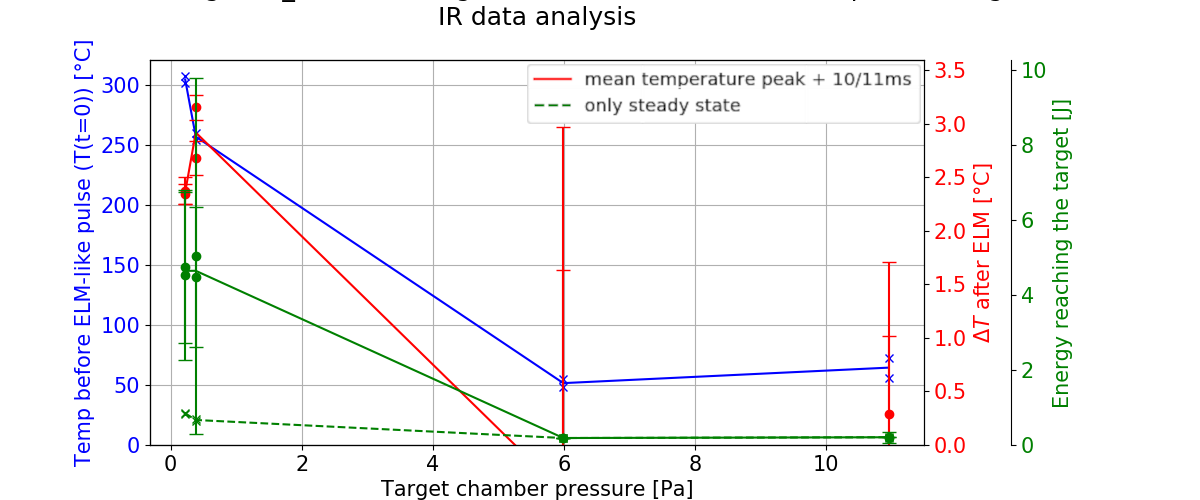}
	\caption{Results of analysis of the IR data for the variations in target chamber neutral pressure. In this figure we focus on the case of weak pulses (10.3J at the source) and lower magnetic field that yields lower plasma pressure background plasma (ID 1 to 4 in \autoref{tab:table1}). Line colors and styles  are the same as \autoref{fig:IR2}.}
	\label{fig:IR3}
\end{figure}

In \autoref{fig:IR2}, for strong pulses, and \autoref{fig:IR3}, for weak ones, are shown the results from the analysis of the target temperature. In blue is indicated the surface temperature prior to the ELM-like pulse, which decreases for increasing target chamber neutral pressure as would be expected for increasing volumetric losses in the target chamber and increasing level of detachment in steady state. In red is indicated the average temperature increase 10 to 11ms after the peak. This is done to mitigate the uncertainty in the time of the real temperature peak, hidden by the prompt emission for high neutral pressure cases. This indicates that at low neutral pressure a large portion of the energy of the ELM-like is absorbed by the target, while it is increasingly dissipated in the volume for increasing target chamber neutral pressure. 

In green are indicated the results from the analytical model. In solid green is indicated the total energy absorbed by the target calculated as mentioned above. In Fig. \ref{fig:IR2} the delivered energy decreases to an apparent minimum, possibly due to some energy not delivered via plasma convection but via neutral heating and radiation, shown \cite{Federici2023c} to increase with higher pressure. For the 4.37Pa target chamber pressure case (marked as Stage 1/2 in \autoref{tab:table1}, ID7) the steady state heating is significantly reduced, while the time dependent component remains similar to the lower pressure cases. This suggests that under these conditions, even when the steady state plasma is close to completely detaching from the target, the ELM-like pulse can still reattach dissipating only a small amount of additional energy compared to the lower pressure cases. This is consistent with visible light measurements, indicating increased but still low emission in the bulk of the target chamber. Unfortunately, TS measurements are not available for this case, preventing a full analysis.

In the extreme case of high target chamber neutral pressure and weak ELM-like pulse shown in \autoref{fig:IR3} the energy delivered to the target is negligible, indicating that the pulse energy was mostly dissipated by the gas in the target chamber. In dashed green is indicated the steady state component, always relatively small. The energy delivered decreases with fuelling in a similar fashion as the temperature increase after the pulse. Most notably the energy delivered is negligible for the case with higher neutral pressure.

The thermal analysis allows us to compare the heat fluxes due to an ELM-like pulse in Magnum-PSI with Tokamaks. A metric often used in Tokamaks to estimate the effect of ELMs is the heat flux factor, defined as the energy of the ELM divided by the area in which it is delivered divided by the square root of its duration. This metric is derived by the same analytic solution for the temperature increase in Model1 and is proportional to the maximum temperature increase. The maximum heat flux factor to prevent tungsten melting is $\sim$50MJ$m^{-2} s^{-0.5}$, while $\sim$40MJ$m^{-2} s^{-0.5}$ for the molybdenum target used here. The limit to prevent cracking is about half of the melting one\cite{Pintsuk2007} with some studies indicating it as low as 6MJ$m^{-2} s^{-0.5}$ for tungsten.\cite{Linke2019}

The heat flux factor obtained in these experiments was estimated from the ELM-like pulse energy density from the IR analysis and an arbitrary 0.7ms and 0.3ms pulse duration for the strong and weak pulses respectively (estimated from the fast camera records). In the strong pulses cases, the heat flux factor is decreased from $\sim$2.5MJ$m^{-2} s^{-0.5}$ for p$<$6Pa to $\sim$1.5MJ$m^{-2} s^{-0.5}$ for p$>$6Pa. For weak pulses the drop in heat flux factor went from $\sim$0.4MJ$m^{-2} s^{-0.5}$ for P$<$3Pa to effectively zero for P$>$3Pa. These heat flux factors are higher than what measured in current tokamaks ($<$1.3 for JET) but much lower than what could be achieved in ITER ({600-1100}MJ$m^{-2} s^{-0.5}$ for unmitigated ELMs).\cite{Jachmich2011,Eich2017}

In a tokamak the connection length from target to midplane is of the order of tens of meters on the outer target and the divertor neutral pressure can reach the Pascal range\cite{Kallenbach2018}. If the divertor is separated from the main chamber by a baffle, meaning there is a physical restriction to the flow of neutrals between the divertor and the main chamber, the plasma can plug the divertor entrance and the neutrals cannot freely move between the two. The neutrals have to interact with the plasma in the SOL and (in deep detachment) the ionisation front, and the neutral pressure behind the ionisation front can be increased by more than one order of magnitude compared to that in the main chamber.\cite{Galassi2020,Pitcher2000,Niemczewski1997} In our experiments the connection length in the target chamber is 0.38m and the neutral pressure up to 15Pa. Given the energy delivered to the target by the ELM-like pulse is reduced with neutral pressure, and that tokamaks are characterised by potentially longer connection lengths but similar or higher neutral pressures behind the ionisation front, this result implies that a significant reduction of the heat flux factor should be possible thanks to the cold regions forming close to the target when strongly detached in tokamaks too. The reduction of heat flux factor and heat flux to the target with pressure and connection length could be non linear (see the higher pressure cases in \autoref{fig:IR2}) but \autoref{fig:IR3} shows that transitioning from stage 2 to 3 via volumetric losses should ultimately be possible. The viability of this method to dissipate the ELM energy will ultimately depend on the neutral pressure required to achieve the needed buffering and its effect on the inter-ELM plasma. Configurations like the super-x divertor could be the most suitable for this approach, as the ionisation front could be maintained relatively close to the x-point while a large volume with high neutral pressure can be maintained in the super-x divertor chamber. Reducing the ELM heat flux factor would improve the safety for long term operations in a tokamak, further showing the benefit of baffling  and increasing the neutral chamber pressure in detached scenarios.

In summary, by increasing the target chamber neutral pressure it is possible to cause detachment of the steady state plasma from the target surface. From the visible fast camera brightness in Magnum-PSI, we identify 3 stages for the interaction between the ELM-like pulse and the background gas in the target chamber. By increasing the neutral pressure more interactions happen in the volume rather than in contact with the target, and become more homogeneous in the plasma column. This is true up to Stage 3, where the loss of plasma in the volume becomes dominant. A regime with low-energy ELM-like pulses was found where the gas in the target chamber can effectively prevent all of the pulse energy from reaching the target.

These results are obtained through analysis of one diagnostic at a time. For a deeper understanding, OES, TS, and power source data will be combined within a Bayesian analysis framework, as presented in the second paper associated to this study.\cite{Federici2023c} This will allow us to identify the most significant processes contributing to the increase of energy removed in the volume here demonstrated, and possibly their temporal evolution.

\section{Summary}\label{Summary}

The effect of ELM-like pulses on a detached target in Magnum-PSI was studied with the help of various diagnostics. It was found that the energy of the ELM-like pulse could be effectively dissipated through a high level of detachment, before the ELM energy reaches the target. The decrease in power reaching the target is inferred to be due to higher volumetric power losses in the volume between target and source. The volumetric power losses, inferred from the visible light emission, remain fairly constant along the magnetic field for low neutral pressure. They exhibit a strong peak close to the target due to plasma surface interactions, indicating an attached plasma. As neutral pressure is increased, the volumetric power losses become more uneven along field lines, with the region close to the target decreasing its brightness and expanding towards the plasma source. This change is correlated with an increased loss of plasma before reaching the measuring location. This behaviour can be divided in stages. In Stage 1 the plasma is attached to the target before and during the ELM-like pulse and the energy losses in the volume are at a minimum and are predominantly not radiative. In Stage 2, the target chamber neutral pressure is such that the plasma is detached from the target in steady state but can reattach during the ELM-like pulse, and energy losses in the volume increase. Detachment during the steady state is correlated with a strong increase of the visible hydrogenic line emission, similar to what is inferred from simulations.\cite{Zhou2022} In Stage 3 the plasma is detached also during the pulse and the losses in the volume are such that the plasma cannot reach the target.

The power delivered to the target by the ELM-like pulse is independently calculated by measuring its temperature via an infrared camera. This data set shows a reduction of the power to the target in agreement with the progress in stages mentioned above. In the most extreme cases the power is not detectable, meaning the energy of the ELM-like pulse is completely dissipated in the volume. For the neutral pressure scan with strong pulses the heat flux factor, a metric often used to qualify the damage to the target given by ELMs in tokamaks, is decreased by about half with increasing neutral pressure, becoming negligible at high pressure for weak pulses. This effect could be used in tokamaks, as they can have longer connection length between ionisation front and target, and potentially similar neutral pressures, to reduce the risk of ELMs damaging the target.

In the second paper associated with this study \cite{Federici2023c}, data from OES, TS, and the power supply system will be combined within a Bayesian framework to gain insights into the processes responsible for the observed increase in volumetric power losses.

\begin{acknowledgments}

The authors would like to thank: T. Morgan, Y. Li for the useful discussions on the thermographic analysis; 
H. J. van der Meiden and J. Scholten for providing the Thomson scattering and ADC data.

This work is supported by US Department of Energy awards DE-AC05-00OR22725 and
DESC0014264 and under the auspices of the Engineering and Physical Sciences Research Council [EP/L01663X/1 and EP/W006839/1]. To obtain further information on the data and models underlying this paper please contact PublicationsManager@ukaea.uk.

Support for M. L. Reinke’s contributions was in part provided by Commonwealth Fusion Systems.

This work has been carried out within the framework of the EUROfusion Consortium, funded by the European Union via the Euratom Research and Training Programme (Grant Agreement No 101052200-EUROfusion). Views and opinions expressed are, however, those of the author(s) only and do not necessarily reflect those of the European Union or the European Commission. Neither the European Union nor the European Commission can be held responsible for them.

This work was supported in part by the DIFFER institute.

\end{acknowledgments}

\section{References}
\bibliographystyle{IEEEtran}
\bibliography{references}

\appendix

\section{Sampling strategy}\label{Sampling strategy}

\begin{figure*}
	\centering
	\includegraphics[width=\linewidth,trim={0 0 0 0},clip]{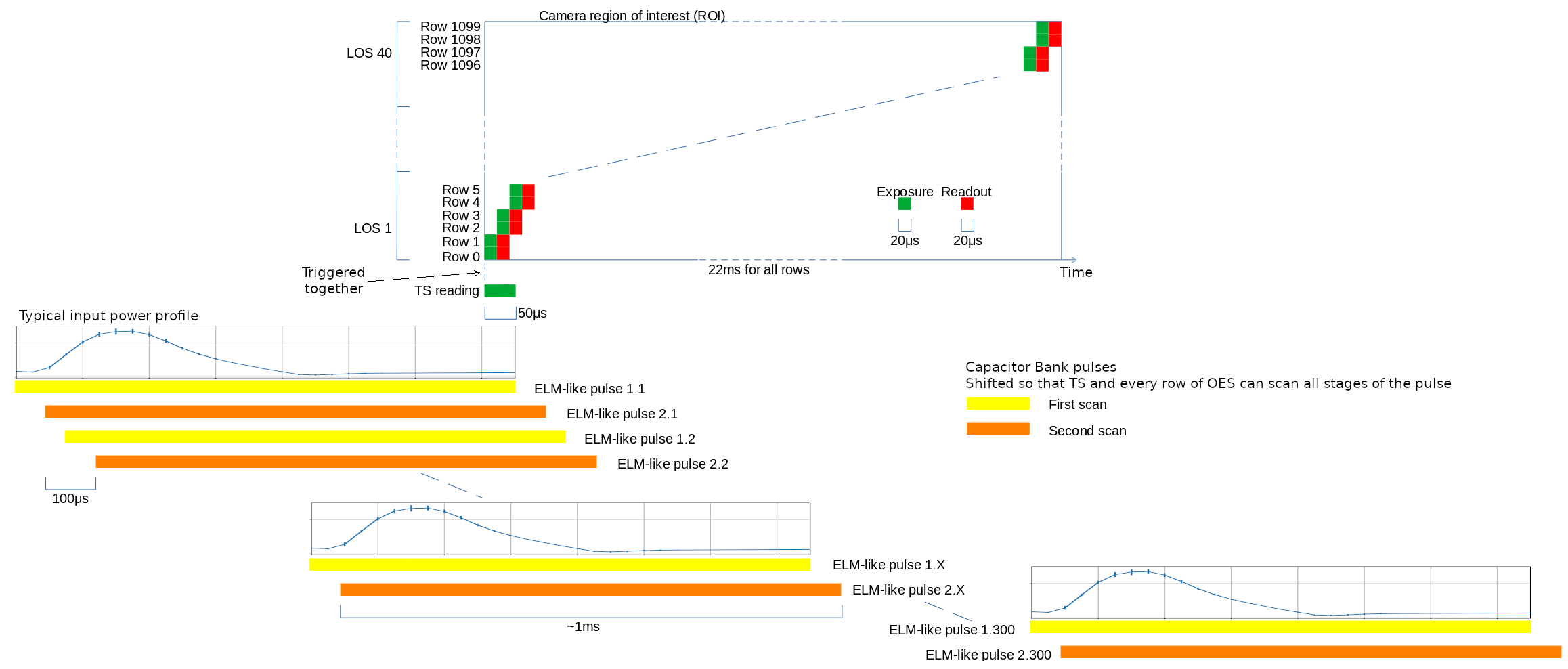}
	\caption{Sampling strategy for TS and OES. The progressive reading of the camera rows is indicated at the top, with the integration time of TS is indicated. Below are each of the ELM-like. pulses. It is shown how the trigger of the capacitor bank is shifted in time and all the pulses are split in two scans. Dashed lines indicate an interruption of the space or time scale.}
	\label{fig:sampling1}
\end{figure*}

As mentioned above, of all diagnostics only the fast camera has a time resolution high enough to resolve individual ELM-like pulses. The TS laser is fired at a fixed frequency of 10Hz from a dedicated timing system. The CB can be triggered at an arbitrary time compared to the 10Hz clock so that information on different parts of the pulse can be collected. The OES is triggered with the same 10Hz signal as TS and, due to the rolling shutter, acquires data with a time shift between rows pairs of $20\mu s$ for a total of $22ms$ required for exposure and acquisition of a single frame. For all experimental conditions, it was decided to record TS/OES data from $0.5ms$ before the ELM-like pulse to $2.5ms$ after. A desired time resolution of $50\mu s$ determines that $(22+3)/0.05=500$ ELM-like pulses, rounded to 600, are required. A phenomena often observed, and only recently solved, was that one capacitor failed to be triggered at the requested time, instead being trigger together with the next. This means that, rather than 600 identical ELM-like pulses, one would be missing, and the one after would be with a released energy twice the others. To reduce the effect of the missing data, the required 600 pulses are split into two 300 pulses scans, with the CB delayed of $100\mu s$ at the time. The two scans are shifted by $50\mu s$ to obtain the desired $50\mu s$ resolution for all the OES rows. With this strategy the maximum time separation between two ELM-like pulses with good data is $100\mu s$ rather than 150$\mu$s. The missing data is obtained by interpolating between the good data. In \autoref{fig:sampling1}, the sampling strategy adopted is represented. At the top is indicated how the camera exposure and readout are shifted in time due to the rolling shutter. 

The rows are managed two at the time, and during the readout of the rows $i-1/i$ the rows $i+1/i+2$ are exposed. Each OES LOS is composed of $\sim 24$ rows, so the time shift within would be $240\mu s$, much larger than the desired $50 \mu s$. Just below is indicated the TS integration time. OES and TS are synchronised such that the trigger to start data acquisition is sent to the two diagnostics simultaneously.
The time difference between CB and OES/TS triggers is initially such that OES and TS reading correspond to the end of the ELM-like pulse. From this, the CB trigger is progressively delayed by $100\mu s$, so that TS and OES can acquire data about progressively earlier stages of the pulse.
For TS this sampling procedure returns directly $T_e$ and $n_e$, while for OES additional steps are required to: reconstruct frames with row data corresponding to the same time slice, bin the LOS, obtain the emission line brightness, calculate radial emissivity from the line integrates brightness. Further details are given in Appendix A of \cite{Federici2023c}. The final result is a map of the line emissivity in both space (radius) and time. The plasma is assumed poloidally symmetric with a radial coordinate of interval 1.06mm dictated by the OES resolution.

\section{Target temperature profile interpretation}\label{Target temperature profile interpretation}
In this section the mathematical models used in the interpretation of the target temperature data will be examined in more detail.

The heat delivered by the plasma to the target during an ELM-like pulse is shaped in time and space. The power density is spatially peaked at the centre of the plasma beam due to the peak in plasma density and temperature. It temporally follows the power evolution dictated by the discharge of the capacitor bank, as measured by the plasma source (see \autoref{fig:TS1} for an example). The full spatial and temporal power density distribution of the target heat source is difficult to obtain from the surface temperature for all experimental conditions, but analytical solutions that can approximate the peak temperature are available. As mentioned in \autoref{IR camera}, the ones considered in this paper are:

\begin{itemize}
\item[Model1] a temporally square heat wave delivered uniformly on a semi infinite plane: the surface temperature increase in the heating and cooling phases are respectively\cite{Behrisch1980}
\begin{equation}
\label{eq:square1}
\begin{aligned}
{\Delta T(t)}_r &= F_0 \frac{2}{ \sqrt{\pi \rho c_p k }} \lbrace { \sqrt{t-t_0} } \rbrace = \frac{E_0}{\tau} \frac{2}{ \sqrt{\pi \rho c_p k }} \lbrace { \sqrt{t-t_0} } \rbrace \\ {\Delta T(t)}_c &= F_0 \frac{2}{ \sqrt{\pi \rho c_p k} } \lbrace { \sqrt{t-t_0} - \sqrt{t-t_0 - {\tau}} } \rbrace
\end{aligned}
\end{equation}
with $\tau$ the duration of the heat pulse and $t_0$ its start, $F_0$ the peak power density, $E_0$ the energy density, $\rho$ the density, $c_p$ the specific heat capacity and $k$ the thermal conductivity
\item[Model2] a temporally square heat wave delivered by a spatially Gaussian beam: the peak surface temperature evolution in the heating and cooling phases are respectively\cite{Bauerle2011}
\begin{equation}
\label{eq:square2}
\begin{aligned}
{\Delta T(t)}_r &= \frac{2}{\pi} {\Theta}_c \tan^{ -1}(2 \sqrt{ \tilde{t}} ) 
\\ 
{\Delta T(t)}_c &= \frac{2}{\pi} {\Theta}_c \tan^{ -1} \left( \frac{2 \sqrt{ \tilde{t}} - 2 \sqrt{ \tilde{t} - \tilde{\tau} }}{ 1+4 \sqrt{ \tilde{t}} \sqrt{ \tilde{t} - \tilde{\tau} }} \right)
\end{aligned}
\end{equation}
with
\begin{equation}
\label{eq:square3}
\begin{aligned}
{\Theta}_c = \frac{\sqrt{\pi}}{2} \frac{F_0 {\omega}_0}{k} , \tilde{t} = \frac{(t-t_0) D}{{\omega}_0} , \tilde{\tau} = \frac{\tau D}{{\omega}_0} , D = \frac{k}{ \rho c_p }
\end{aligned}
\end{equation}
and $\omega_0$ the $1/e$ size of the spatial heat distribution on the target
\item[Model3] a temporally triangular heat wave delivered by a spatially Gaussian beam: the peak surface temperature evolution in the heat rise, heat decreasing and cooling phases are respectively

\begin{equation}
\label{eq:triang}
\begin{aligned}
{\Delta T(t)}_r &= \frac{2}{\pi} {\Theta}_c \frac{1}{\tilde{\tau_r}}   \left\{ {\tilde{t}} \tan^{ -1}(2 \sqrt{ \tilde{t}} )  - \frac{1}{4} [ 2 \sqrt{ \tilde{t}} - \tan^{ -1}(2 \sqrt{ \tilde{t}} ) ] \right\}
\\ 
{\Delta T(t)}_d &= {\Delta T(t)}_r - \frac{2}{\pi} {\Theta}_c \left( \frac{1}{\tilde{\tau_r}} +\frac{1}{\tilde{\tau_d}}  \right) \left\{ {(\tilde{t} - \tilde{\tau_r})} \tan^{ -1} (2 \sqrt{ \tilde{t} - \tilde{\tau_r}} )  \right. 
\\
& \phantom{+} \left. - \frac{1}{4 } [ 2 \sqrt{ \tilde{t} - \tilde{\tau_r}} - \tan^{ -1}(2 \sqrt{ \tilde{t} - \tilde{\tau_r}} ) ] \right\} 
\\ 
{\Delta T(t)}_c &= {\Delta T(t)}_d + \frac{2}{\pi} {\Theta}_c \frac{1}{\tilde{\tau_d}} \left\{ {(\tilde{t} - \tilde{\tau_r} - \tilde{\tau_d})} \tan^{ -1}(2 \sqrt{ \tilde{t} - \tilde{\tau_r} - \tilde{\tau_d}} ) \right. 
\\
& \phantom{+} \left. - \frac{1}{4 } [ 2 \sqrt{ \tilde{t} - \tilde{\tau_r} - \tilde{\tau_d}} - \tan^{ -1}(2 \sqrt{ \tilde{t} - \tilde{\tau_r} - \tilde{\tau_d}} ) ] \right\}
\end{aligned}
\end{equation}
with $\tilde{\tau_r}$ the time to rise the power density from $0$ to $F_0$ and $\tilde{\tau_d}$ the time to decrease to $0$ again.

\end{itemize}
These have been obtained assuming homogeneous and constant thermal properties and a semi infinite target. With these assumptions the heat equation is linear and the superposition principle could be employed.

An important observation from these analytical solutions is that in all cases, in the limit $t>>\tau$ the surface temperature cooling is proportional to $t^{-3/2}$. For pulses with a significant energy delivered, the target surface is still cooling when the next ELM-like pulse comes, so the temperature has to be corrected. The temperature before the pulse is fited with
\begin{equation}
\label{eq:squareSS}
\begin{aligned}
T=\frac{a}{ (t+ \Delta t)^{3/2}} + T_{0}
\end{aligned}
\end{equation}
with $\Delta t$ the time between consecutive ELM-like pulses. The time dependent component is then subtracted.

\begin{figure}
	\centering
	\includegraphics[width=\linewidth,trim={750 550 10 115},clip]{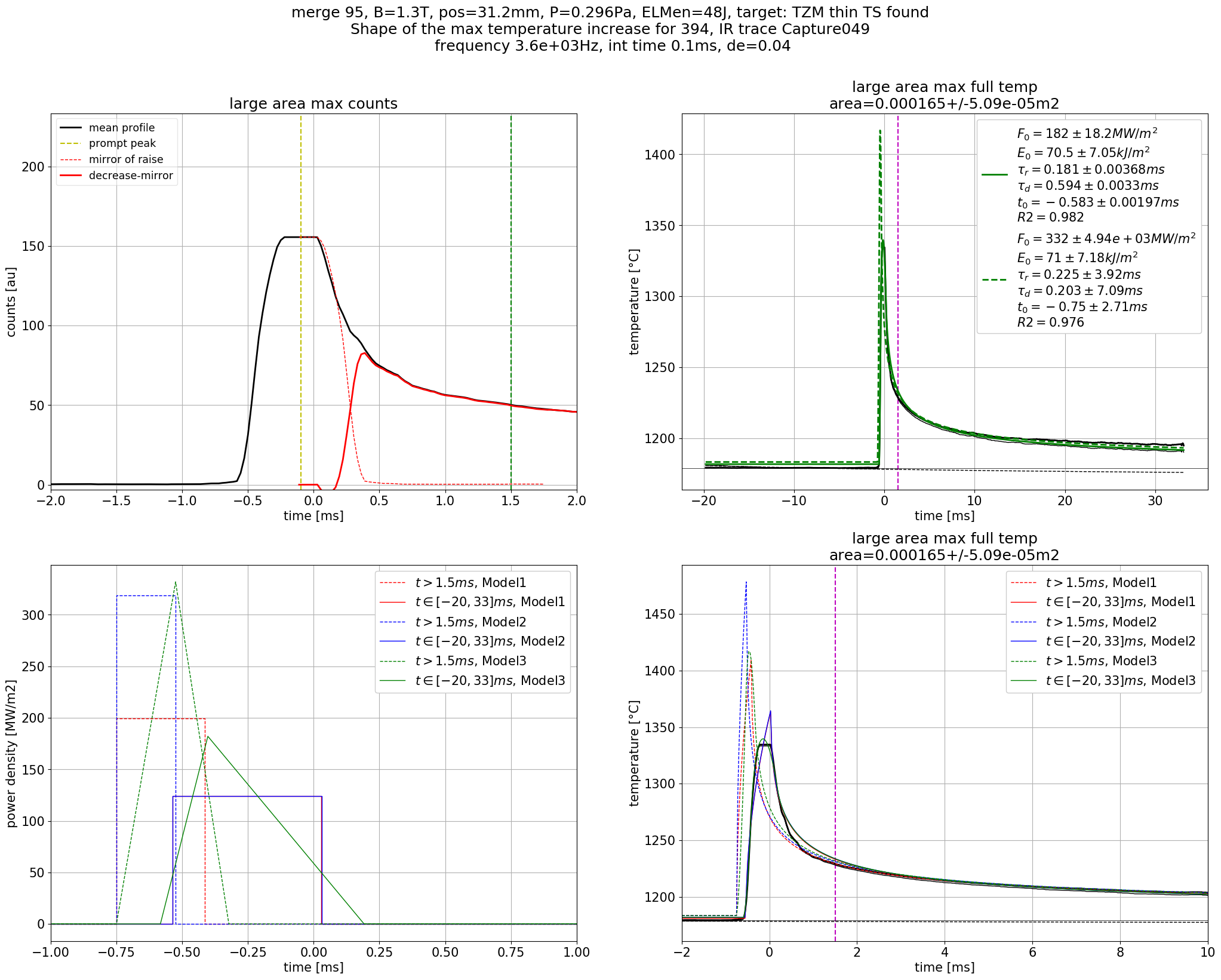}
	\caption{Measured and fitted peak target temperature for a discharge in Stage 1 (ID 5 in \autoref{tab:table1}). In dashed green the fit of the temperature profile using the triangular Gaussian pulse model for t$>$1.5ms and in solid green the whole profile. The black dashed line indicates the cooling from the previous pulse and the thin solid black one is the peak target temperature not corrected for this.}
	\label{fig:IR4}
\end{figure}

\begin{figure}
	\centering
	\includegraphics[width=\linewidth,trim={5 5 55 65},clip]{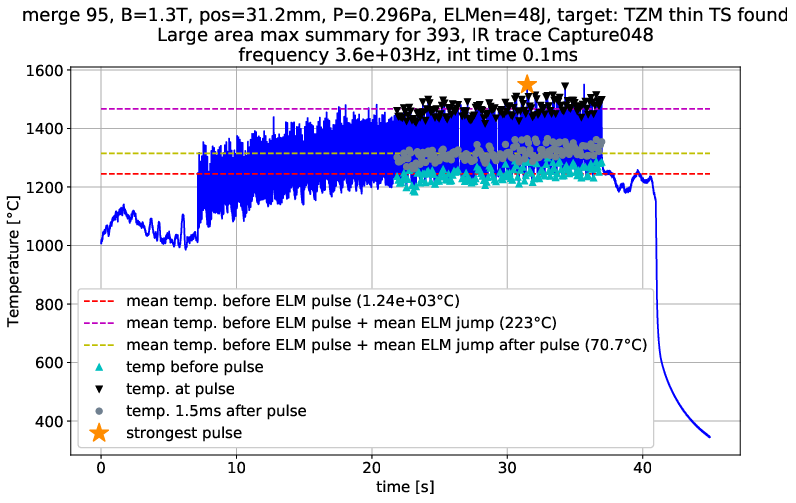}
	\caption{Peak temperature of the target for low neutral pressure (ID 5 in \autoref{tab:table1}). To estimate the effect of the ELM-like pulse only the second half of the pulses is used, when the target is close to a steady state.}
	\label{fig:IR5}
\end{figure}

The result can be seen in the black dashed line in \autoref{fig:IR4} and in the increase of the peak temperature from the thin solid black line to the thick one. This correction is significant only for low neutral pressure conditions. In these cases it has to be taken also into account that it takes time to reach a thermal equilibrium after the ELM-like pulse train is started. To minimize this slow variation only the second 150 of the ELM-like pulses within a 300 strong scan is used to construct the average profile as it can be seen from \autoref{fig:IR5}.

\begin{figure}
    \hspace*{+10mm}
     \centering
     \begin{subfigure}{0.7\linewidth}
         \centering
         \includegraphics[width=\textwidth,trim={510 5 5 415},clip]{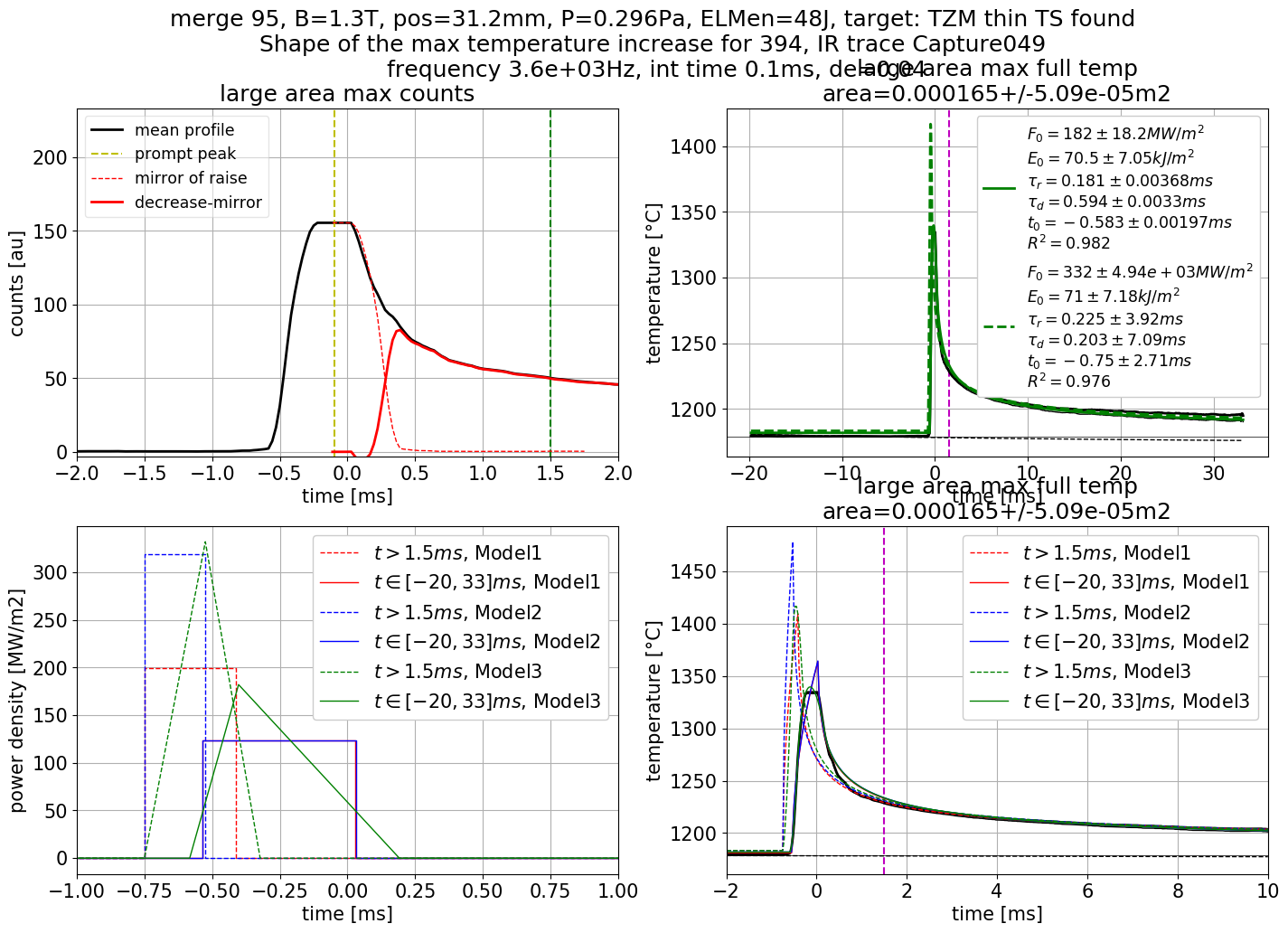}
         \vspace*{-30mm}
         \caption{\phantom{ww}}
         \label{fig:IR6a}
     \end{subfigure}
     \hfill
    \vspace*{+22mm}
    \begin{subfigure}{0.7\linewidth}
         \centering
         \includegraphics[width=\textwidth,trim={5 5 515 415},clip]{chapter3/figs/file_index_394_IR_trace_Capture049_43.png}
         \vspace*{-30mm}
         \caption{\phantom{wew}}
         \label{fig:IR6b}
     \end{subfigure}
        \vspace*{+20mm}
        \caption{Comparison of the peak power density profiles (\subref{fig:IR6b}) result of fitting the peak temperature curve (\subref{fig:IR6a}) the with different analytical solutions for the lowest available neutral pressure conditions (ID 5 in \autoref{tab:table1}). All find adequate values for the energy density even with radically different peak durations. Dashed lines are obtained fitting from 1.5 to 30ms while solid ones from -20 to 30ms. In magenta 1.5ms after the temperature peak.}
        \label{fig:IR6}
\end{figure}

Considering all this, a fit of the ELM-like pulse with the lowest neutral pressure can be used to compare the performance of the different models. As shown in \autoref{fig:IR6} all models fit well the slowly decreasing peak temperature slope and return a power density within 10\% of each other. Model3 fits best also the peak when the entire time axis is used.

\begin{figure}
    \hspace*{+10mm}
     \centering
     \begin{subfigure}{0.7\linewidth}
         \centering
         \includegraphics[width=\textwidth,trim={510 5 5 330},clip]{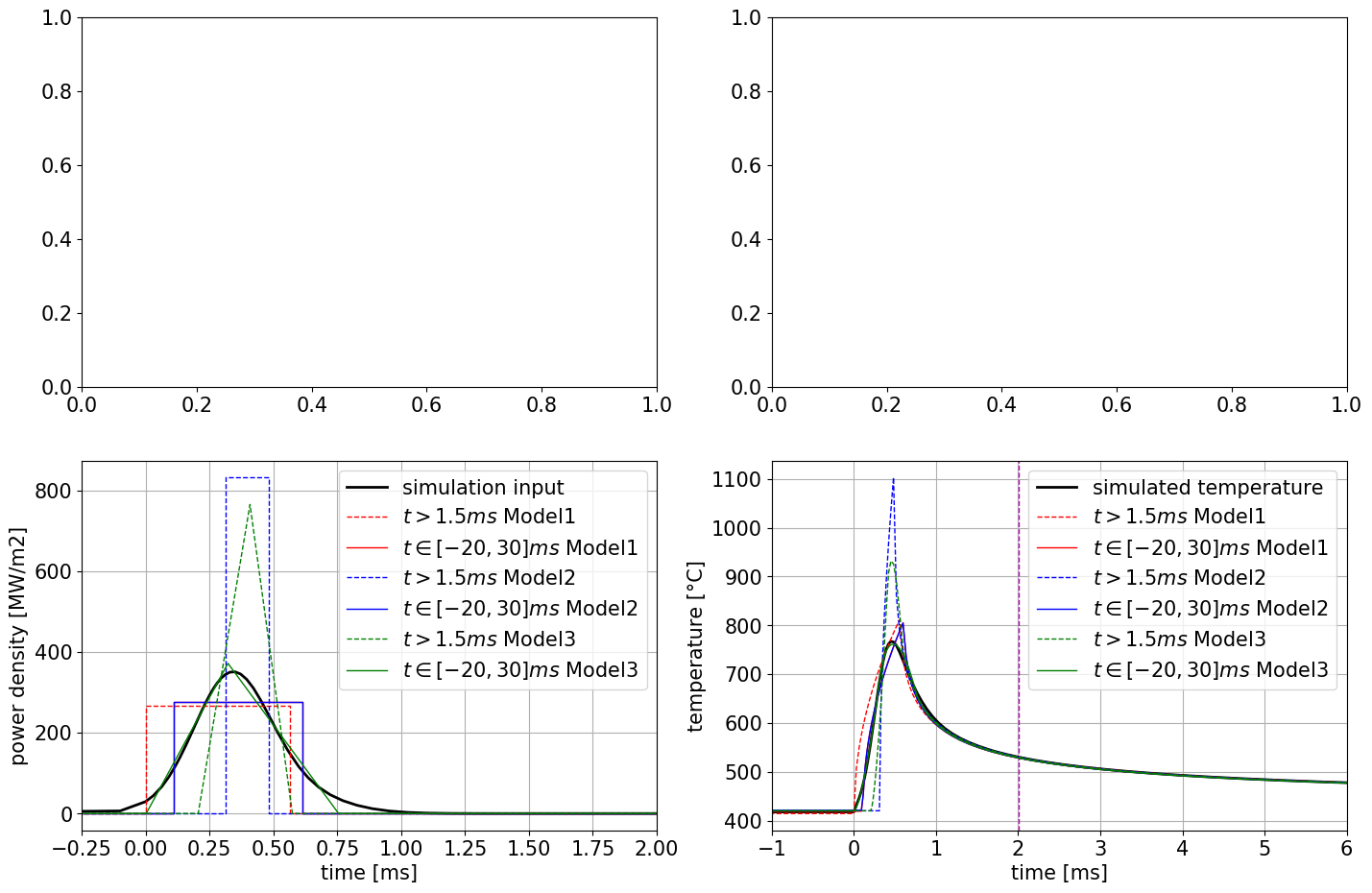}
         \vspace*{-30mm}
         \caption{\phantom{wew}}
         \label{fig:IR7a}
     \end{subfigure}
     \hfill
    \vspace*{+22mm}
     \begin{subfigure}{0.7\linewidth}
         \centering
         \includegraphics[width=\textwidth,trim={5 5 515 330},clip]{chapter3/figs/Figure_3.png}
         \vspace*{-30mm}
         \caption{\phantom{wew}}
         \label{fig:IR7b}
     \end{subfigure}
        \vspace*{+20mm}
        \caption{Peak target temperature evolution (\subref{fig:IR7a}) calculated with the MSC.Marc/Mentat® code from a known heat flux profile made to be similar to the one caused by the CB in Magnum-PSI (\subref{fig:IR7b}) (solid black lines). That is fit with different analytical solutions and the inferred heat flux compared to the input. Dashed lines are obtained fitting from 1.5 to 30ms while solid ones the entire time axis. In magenta 1.5ms after the temperature peak.}
        \label{fig:IR7}
\end{figure}

The same comparison can be done with a simulated temperature rise obtained from a known input power profile. The MSC.Marc/Mentat® non linear FEM suite was used to reproduce a heat pulse similar to the one generated by the CB in terms of temporal and spatial variation and with temperature dependent material properties. The spatial distribution of the heat pulse was set as a Gaussian of radial extent consistent with what measured with TS ($1/e^2 \sim 1$cm) while the temporal variation was found by fitting two Gaussians to the power profile from the plasma source. The comparison of the fits using the 3 models is shown in \autoref{fig:IR7}.

Model3 is capable of better reproducing the full temperature profile returning, for the full fit, with a power density shape very close to the input one. Even if the pulse duration when fitting only after 1.5ms is quite different from the input, the energy density is within 10\% of the input one for all cases and, as mentioned before, the uncertainty on the energy delivered to the target is $\sim$20\%, good enough for the purpose of this paper.

\begin{figure}
    \captionsetup{labelfont={color=black}}
     \centering
     \begin{subfigure}{1\linewidth}
         \centering
         \vspace*{-0mm}
         \includegraphics[width=\textwidth,trim={0 0 10 60},clip]{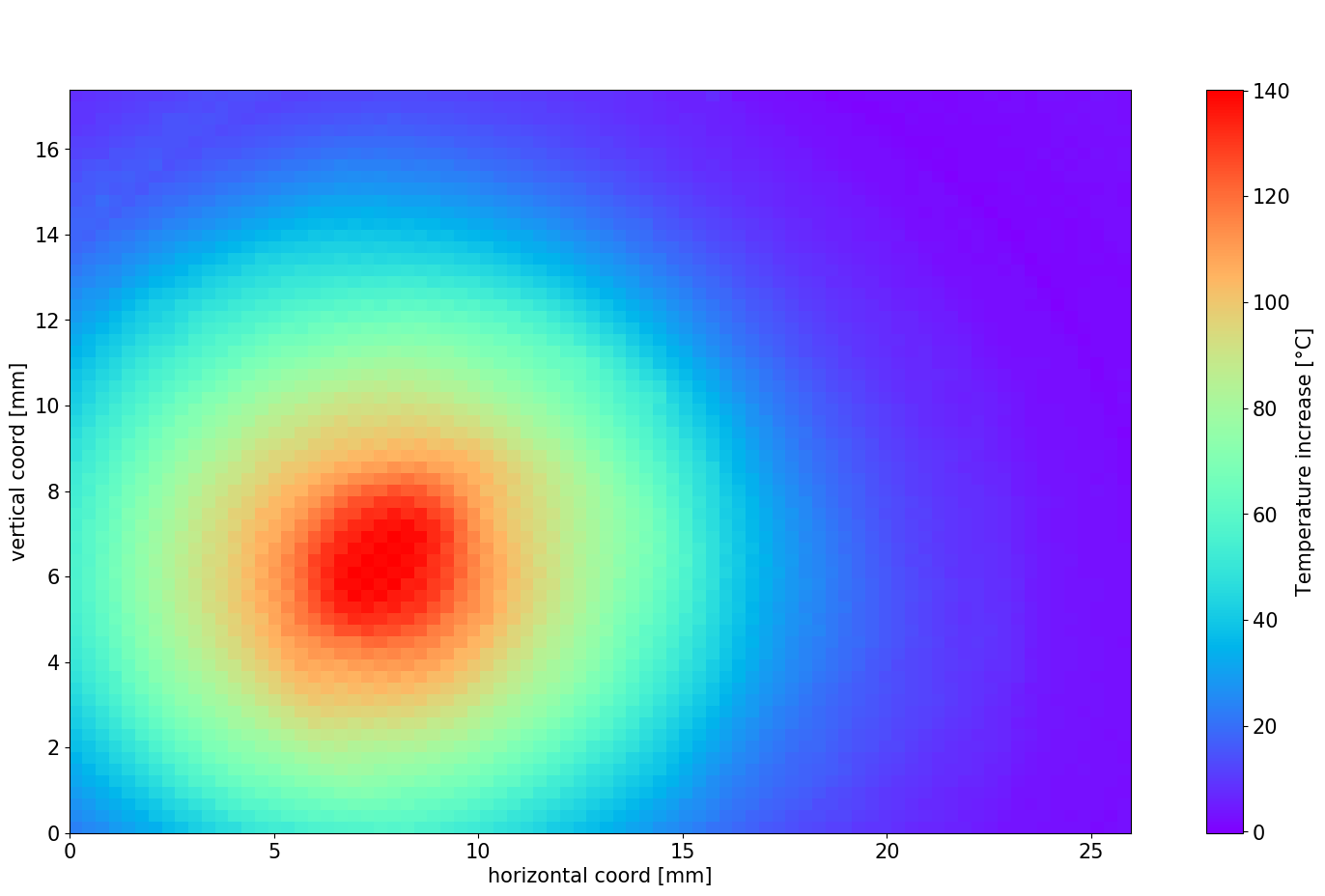}
        \vspace*{-30mm}
        {\color{white}\caption{\phantom{wewew}}\label{fig:IR8a}}
     \end{subfigure}
     \hfill
    \vspace*{+22mm}
     \begin{subfigure}{1\linewidth}
         \centering
         \vspace*{-0mm}
         \includegraphics[width=\textwidth,trim={0 0 10 66},clip]{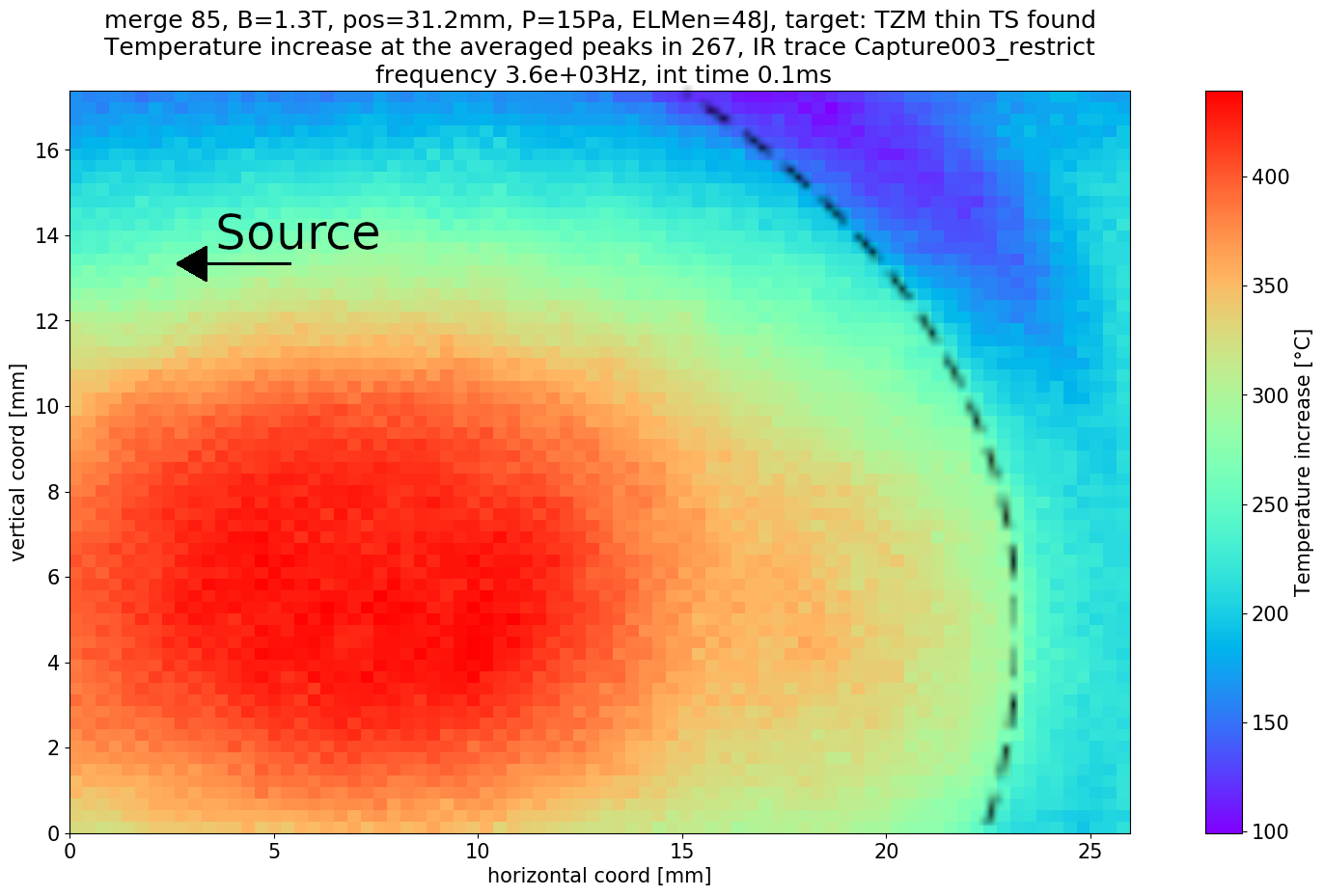}
        \vspace*{-30mm}
         {\color{white}\caption{\phantom{}}\label{fig:IR8b}}
     \end{subfigure}
        \vspace*{+20mm}
        \caption{Peak temperature increase distribution over the steady state on the target. (\subref{fig:IR8a}) low neutral pressure conditions showing a clear confined peak (ID 5 in \autoref{tab:table1}). (\subref{fig:IR8b}) high neutral pressure case showing the temperature increase is not localized, possibly consistent with reflections of volumetric radiation or radiation of other origin (ID 10 in \autoref{tab:table1}). The black dashed line is the circular side of the cylindrical target. The arrow indicates where the source is outside the camera view.}
        \label{fig:IR8}
\end{figure}

To reinforce the argument on the origin of the prompt emission for high neutral pressure cases, the temperature profile at its peak for a low and high neutral pressure case is shown in \autoref{fig:IR8}. The low neutral pressure case (\subref{fig:IR8a}) shows a well defined peak with a radially symmetric decreasing profile. The high neutral pressure case (\subref{fig:IR8b}) is very different, with a much wider peak and a high temperature shadow from the peak to the edge of the target (marked in black). Note also the higher temperature outside the target itself. Considering the camera looks at the target at an angle and that the source is at the left of the field of view (see \autoref{fig:layout}), it is possible that the prompt peak could be due to radiation from the plasma itself reflected by the target. A definitive answer on what the origin of this radiation cannot be given at this stage, but this observation further strengthen the case for only fitting the slowly decreasing temperature profile after the pulse.

\end{document}